\documentclass[5p,twocolumn]{elsarticle}

\usepackage[utf8]{inputenc}
\usepackage{lineno}
\usepackage[hidelinks]{hyperref}
\hypersetup{colorlinks=true}
\usepackage[english]{babel}
\usepackage{caption}			
\usepackage{xspace}				
\usepackage{amsmath}			
\usepackage{amssymb}	
\usepackage{graphicx}
\usepackage{booktabs}				
\usepackage{units}					
\usepackage{multirow}				
\usepackage{tabularx}				
\usepackage{miller}					
\usepackage{siunitx}				
\usepackage{pdfpages}

\newcommand{\etal}{\textit{et al.}}

\newcommand{\degree}{\textordmasculine}
\newcommand{\degrees}{\textordmasculine\enspace}

\newcommand{\MgAl}{\textnormal{Mg}$_{17}$\textnormal{Al}$_{12}$}
\newcommand{\tauC}{$\tau_{c}$\xspace}
\journal{Materialia}

\biboptions{numbers, sort&compress}

\begin{document}

\begin{frontmatter}
	
	\title{Atomistic Simulations of Basal Dislocations Interacting with\\ \MgAl{}  Precipitates in Mg}
	
	\author[a]{Aviral Vaid}
	\author[b,a]{Julien Gu\'enol\'e \corref{author}}
	\author[c,a]{Aruna Prakash}
	\author[b]{Sandra Korte-Kerzel}
	\author[a]{Erik Bitzek}
	
	\cortext[author] {Corresponding author.\\\textit{E-mail address:} guenole@imm.rwth-aachen.de}
	\address[a]{Department of Materials Science and Engineering, Institute I, Friedrich-Alexander-Universit\"at Erlangen-N\"urnberg (FAU), \\Martensstr. 5, 91058 Erlangen, Germany}
	\address[b]{Institute of Physical Metallurgy and Metal Physics, RWTH Aachen University, Germany}
	\address[c]{Micromechanical Materials Modelling (MiMM), Institute of Mechanics and Fluid Dynamics, \\ Technische Universität Bergakademie Freiberg (TUBAF), Germany}

%
	

\begin{abstract}
The mechanical properties of Mg-Al alloys are greatly influenced by the complex intermetallic phase \MgAl, which is the most dominant precipitate found in this alloy system. 
The interaction of basal edge and 30\degree{} dislocations with \MgAl{} precipitates 
is studied by molecular dynamics and statics simulations, varying the inter-precipitate spacing ($L$), and size ($D$), shape and orientation of the precipitates.
The critical resolved shear stress $\tau_c$ to pass an array of precipitates follows the usual $\ln((\nicefrac{1}{D} + \nicefrac{1}{L})^{-1})$ proportionality.
In all cases but the smallest precipitate, the dislocations pass the obstacles by depositing 
dislocation segments in the disordered  interphase boundary rather than shearing the precipitate or leaving Orowan loops in the matrix around the precipitate. 
An absorbed dislocation increases the stress necessary for a second dislocation to pass the precipitate also by absorbing dislocation segments into the boundary. 
Replacing the precipitate with a void of identical size and shape decreases the critical passing stress and work hardening  contribution while an artificially impenetrable \MgAl{} precipitate increases both. These insights will help improve mesoscale models of hardening by incoherent particles.
%
\end{abstract}
	
	
\end{frontmatter}



\section{Introduction}
\label{Introduction}





Precipitation strengthening 
is known to enhance yield and flow stress of metallic alloys by the presence of second-phase particles distributed in a homogenous matrix that impede the motion of dislocations~\cite{Nembach1997}.
Various factors are known to influence the strengthening behavior in metallic alloys such as the size, shape, number and distribution of precipitates, the crystallographic alignment between the phases, the interfacial energy and the stacking fault and dislocation energies within the precipitates~\cite{Bitzek2005, Bacon2009DIS, Osetsky2003, Shin2003, Takahashi2008, Kohler2005, Terentyev2008, Terentyev2012, Takahashi2010}. 

This strengthening effect is exploited in many engineering alloys, including magnesium. 
Mg alloys have recently attracted significant interest due to their low density, high specific strength, and good recyclability~\cite{ Polmear1994, Luo1999, Hanko2002, Bohlen07MSF}. 
They are of interest  as structural materials for the automotive and aeronautical industries where a strong reduction in components' weight lowers fuel consumption and gas emissions~\cite{Bamberger2008}. 
Understanding the plastic deformation mechanism in such precipitation-strengthened alloys is of prime importance for improving the modeling of light-weight materials and for developing novel alloys with improved properties.
The main alloying element in most Mg alloys is aluminum, which leads to the formation of several intermetallic phases, the most common being 
\MgAl{}~\cite{Wang2008b, Clark1968, Burssik2002, Polmear1994}. 
\MgAl{}, therefore, plays a key role in increasing the yield strength of Mg-Al alloys mainly by hindering the easy glide of dislocations on the basal plane of the Mg-matrix.



The interactions between dislocations and discrete obstacles have been studied since the early days of dislocation theory using experimental techniques~\cite{Humphreys1970, Hirsch1957, Delmas2003, Liu2011a, Takahashi2012}, and linear elasticity~\cite{Nembach1983, Brown1971, Hazzledine1974} as well as, more recently, dislocation dynamics modeling~\cite{Shin2003, Monnet2011, Xiang2004, Queyreau2010, Xiang2006, Takahashi2010} and atomistic simulations~\cite{Osetsky2003, Bitzek2005, Kohler2005, Hatano2006, Takahashi2008a, Terentyev2008, Bacon2009, Proville2010, Terentyev2011, Terentyev2012, Prakash15AM}. 
A detailed review of atomic-scale modeling of dislocation-obstacle interactions can be found in~\cite{Bacon2009DIS}. 
Most simulation studies on dislocation-obstacle interactions have so far been performed on fcc and bcc metals. 
Here, typical obstacles include voids or vacancy clusters and precipitates,
either as artificially impenetrable obstacles or by considering realistic material-specific alloy systems. 
Currently, only comparatively few simulation studies exist in the literature that specifically address dislocation-obstacle interactions in hcp metals, like the recent molecular static simulations by Groh~\cite{Groh2014} who studied basal edge dislocations in Mg interacting with spherical obstacles generated by artificially immobilizing Mg atoms.
Even fewer atomistic simulation studies exist on dislocations in simple 
hcp, fcc or bcc metals interacting with complex intermetallic precipitates ~\cite{fan2018effect, fan2018precipitation, Liao2014}.

%
Regarding the Mg-Al system of interest in this study, most of the 
recent work was performed in the group of Horstemeyer\cite{Liao2014, Liao2013, Liao2013a}. 
However, the properties of the complex intermetallic phase \MgAl{} depend crucially on the used interatomic potential. 
For example, the modified embedded atom model (MEAM) potential by Jelinek \etal{}~\cite{Jelinek2012a} used by Liao \etal{}~\cite{Liao2014, Liao2013, Liao2013a} gives a negative shear modulus $C_{44}$ and a positive enthalpy of formation for the stable \MgAl{} phase.
More recently, Moitra and Llorca~\cite{Moitra2017} used an embedded atom model (EAM) potential instead of an MEAM potential for the Mg-Al system. However,
this potential has a problem with the matrix phase, as EAM potentials, in general, inaccurately represent the elastic constants of hcp materials~\cite{Pasianot1992a}. 
For a detailed analysis and comparison of Mg-Al potentials the reader is referred to the supplementary material section S1.



In the following, we present a detailed qualitative as well as quantitative
atomistic simulation study of edge and mixed basal dislocations in Mg interacting with different types of obstacles, with the focus on \MgAl{} precipitates of varying size, shape and spacing.

\section{Methodology}
\label{sec:Methodology}

The atomistic simulations presented in this work were performed with the classical molecular dynamics code LAMMPS~\cite{Plimpton95JCP} (version \texttt{16 Mar 2018}). 
Atomic interactions were modeled  by the MEAM potential of Kim \etal{}~\cite{Kim2009}, which was found to best represent an atomistic system consisting of both Mg and \MgAl{} (see supplementary material section S1 for a comparison study between different potentials).

\subsection{Sample Setup}

\begin{figure}
	\includegraphics[width=1.0\linewidth]{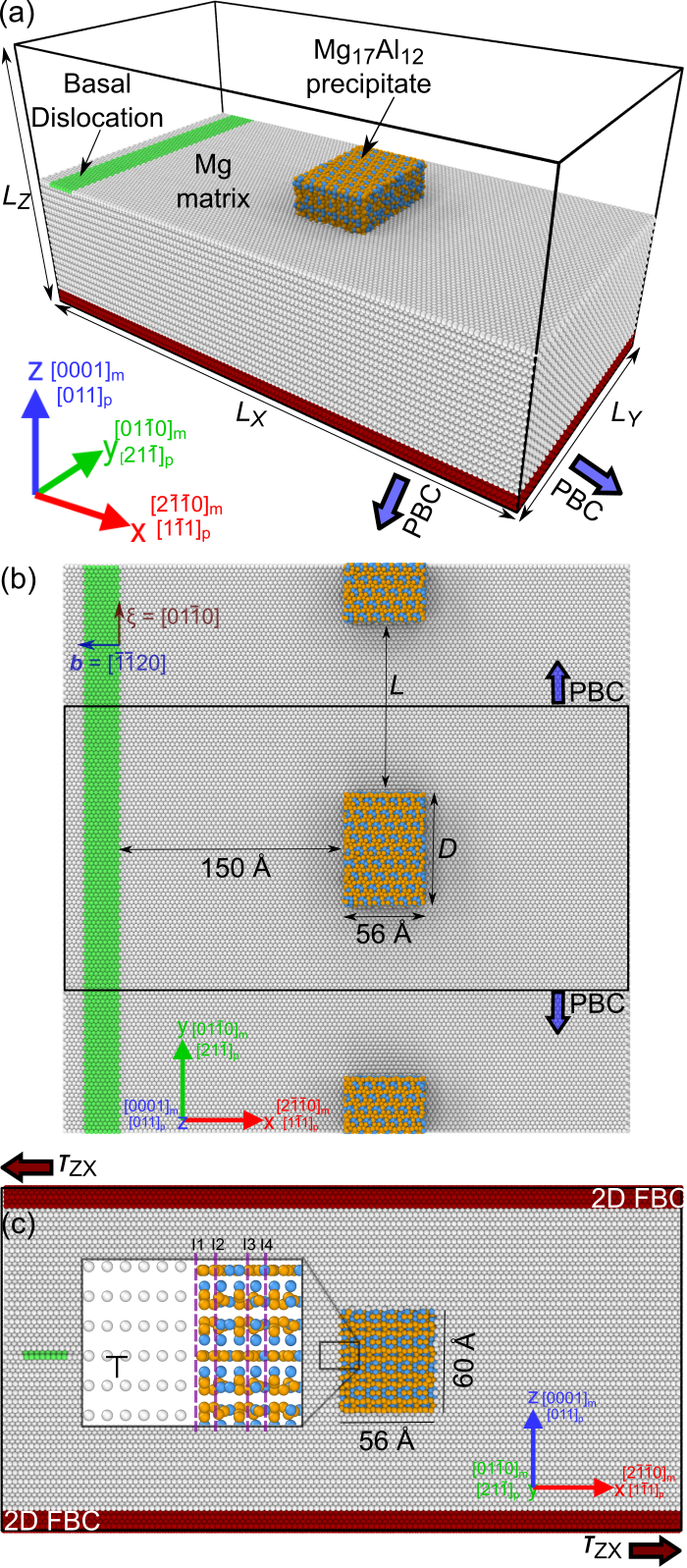}
	\caption{Illustrations of the simulation setup a) perspective and b) top view, with the upper half of  Mg atoms removed for clarity. c) side view, cut in the middle of the Y axis. The inset shows the relative alignment of slip plane in Mg to that in \MgAl{}. The purple lines mark the interface termination (coded as 1, 2, 3, 4). Color code of precipitate atoms based on chemical species: Mg, orange; Al, Blue. Color code of Mg matrix atoms based on local crystallography: fcc, green; hcp, other, grey. Atoms in red are part of the 2D dynamic boundaries. PBC: Periodic boundary conditions, FBC: Force boundary conditions. The simulation box is marked in black.}
	\label{fig:SimulationSetup}
\end{figure}

A typical simulation setup used in the present work is shown in figure \ref{fig:SimulationSetup}. 
It consists of an hcp-Mg matrix with the basal \hkl(0001) plane-normal parallel to the \textit{Z}-axis, a dislocation with line direction \textbf{\textit{$\xi$}} = \hkl[01-10] parallel to the \textit{Y}-axis
and a precipitate. 
Periodic boundary conditions (PBC) were used along the dislocation line direction and the direction of dislocation motion, \textit{i.e.}, along the X and Y directions.
This setup corresponds therefore to an array of infinite dislocations that interacts with a periodic array of obstacles, as illustrated in fig.~\ref{fig:SimulationSetup}(b), and allows the study of multiple interactions of the dislocation with the obstacles.
Atoms in the top and bottom boundary layers were constricted to move only within the Z-plane (2D dynamic boundary layers).

\begin{table*}
	\centering
	\caption{Characteristics of the simulated samples. 
		The samples were varied within a \textit{parameter range} with a \textit{default value} used to build the \textit{default setup}.
		$L$ denotes the inter-precipitate spacing, $\mathbf{b}$ the Burgers vector and $\mathbf{\xi}$ the line direction.
	}
	
	\label{tab:ParametersSpace}
	\small
	\begin{tabularx}{\textwidth}{@{}p{5cm} p{9.5cm} p{3cm}@{}}
		\toprule
		Parameters & Parameter Range & Default setup \\ 
		\midrule
		\textit{Precipitate spacing} \\
		-- $L$ (\AA) & 122, 222, 322, 422, 522 & 122 \\
		\midrule
		\textit{Precipitate geometry} & & \\
		-- shape  &  cuboidal, column, sphere  & cuboidal \\
		-- width, cuboidal  (\AA) & 78, 156, 234 & 78      \\
		-- diameter, sphere (\AA) & 20, 50, 78 & --   \\
		\midrule
		\textit{Matrix-precipitate interface}& & \\
		-- orientation relationship (OR) & Burgers OR($L=\SI{122}{\AA}$, $D=\SI{78}{\AA}$), & \\ 	   	& \hkl[1-11]$_{p}$ $\parallel$ \hkl[2-1-10]$_{m}$ and \hkl(21-1)$_{p}$ $\parallel $\hkl(0001)$_{m}$ ($L=\SI{122}{\AA}$, $D=\SI{78}{\AA}$),  & Burgers OR \\
		& \hkl[011]$_{p}$ $\parallel$ \hkl[2-1-10]$_{m}$ and \hkl(-11-1)$_{p}$ $\parallel $\hkl(0001)$_{m}$ ($L=\SI{122}{\AA}$, $D=\SI{78}{\AA}$)&  \\
		-- slip plane alignment  & \hkl(0001)$_m$/\hkl(011)$_p$, \hkl(0001)$_m$/\hkl(033)$_p$ & \hkl(0001)$_m$/\hkl(011)$_p$ \\
		-- termination (type) & I1, I2, I3, I4 & I1 \\
		\midrule
		\textit{Dislocation} \\
		-- character  & edge (\textbf{\textit{b}} = \nicefrac{a$_0$}{3} \hkl[-2110], \textbf{\textit{$\xi$}} = \hkl[01-10]), & edge \\
		& mixed 30\degree (\textbf{\textit{b}} = \nicefrac{a$_0$}{3} \hkl[-1-120], \textbf{\textit{$\xi$}} = \hkl[01-10]) & \\
		\midrule
		\textit{Obstacle} \\
		-- type	& void, precipitate, impenetrable precipitate & precipitate \\ 
		\bottomrule
	\end{tabularx}
\end{table*}

Edge and 30\degree{} dislocations with Burgers vectors \textbf{\textit{b}} = \nicefrac{$a_0$}{3} \hkl<-2110> were introduced following the method detailed in \cite{Rodney2000} and \cite{Bitzek2005}, where $a_0$ is the lattice constant of Mg at \SI{0}{K}.
The precipitates were inserted approx. 150 \AA{} in front of the dislocation (see fig.~\ref{fig:SimulationSetup}(b)) by removing the matrix atoms and filling the 
resulting void with Mg and Al atoms arranged in the bcc-like crystal structure of the \MgAl{} phase (space group $I\overline{4}3m$)~\cite{Wang2008b, Zhang2005a}.
The relative orientation of the Mg matrix and the \MgAl{} precipitate follows the experimentally-observed Burgers orientation relationship (OR)~\cite{Clark1968}: \hkl[1-11]$_{p}$ $\parallel$ \hkl[2-1-10]$_{m}$ and \hkl(011)$_{p}$ $\parallel $\hkl(0001)$_{m}$.  
The glide plane of the dislocations were aligned with a 
\hkl{011} plane in \MgAl{}, which is assumed to be the natural glide planes of the phase~\cite{Xiao2013, Hagihara2018}, thus presumably allowing for easy slip transmission into the precipitate.

In the present work, we varied the inter-precipitate spacing, precipitate geometry, matrix-precipitate interface, dislocation character and obstacle type (\MgAl{} precipitate, \MgAl{} precipitate with artificially frozen atoms and a void). Due to the complex crystallography of the precipitate, we prepared various interfaces that follow the Burgers OR while modifying the interface termination at the precipitate (see inset fig. \ref{fig:SimulationSetup}(c), supplementary material section S2). The same cuboidal precipitate was additionally rotated by 90\degrees anti-clockwise along X-axis and along Y-axis relative to the position shown in fig. \ref{fig:SimulationSetup}(b) to create setups with different interfaces.
The variations of the simulation setup are summarized in tab.~\ref{tab:ParametersSpace}. 
Most simulations were performed using the values in the last column (tab. ~\ref{tab:ParametersSpace}) and the setup is hereto referred as \textit{default setup}. 
To study the influence of each parameter, only one characteristic was modified at a time relative to the \textit{default setup} while keeping the others constant. Care was taken to minimize the influence of box size and spurious image forces on the quantitative evaluation of CRSS in our simulations following Szajewski \etal~\cite{Szajewski2015} (see supplementary material section S3)

\subsection{Atomistic Simulations}

The created samples were relaxed by first using a local relaxation~\cite{Sheppard2008} followed by a full relaxation using an optimized implementation~\cite{Guenole19FIRE} of the FIRE~\cite{Bitzek2006} algorithm to reach the equilibrium state. 
The samples were considered sufficiently relaxed when the force norm, i.e., the norm of the 3N-dimensional force vector, fell below a threshold value of $10^{-8}$ eV/{\AA}.

The critical resolved shear stress (CRSS) required for a dislocation to overcome the precipitates was computed using a series of molecular statics (MS) simulations. 
Each setup was pre-sheared along $Z$ in direction of the Burgers vector according to the value of the desired resolved shear stress, using the corresponding elastic constants of the Mg matrix (see supplementary material section S4)
Forces corresponding to the desired shear stress were applied to the atoms in the top and bottom layers parallel to the glide plane (2D FBC, see fig \ref{fig:SimulationSetup}). 
This leads to a symmetric shear state on the slip plane situated in the middle of the setup. 
The CRSS required for a dislocation to overcome a precipitates was determined by bracketing. 
The \textit{default} setup was also unloaded from a relaxed configuration by removing any externally applied shear stresses and allowing the system to reach an energy minimized state.

The dynamics of the dislocation-obstacle interactions were furthermore studied using molecular dynamics (MD) simulations, using either the NVE ensemble with an initial temperature of $T_0 = \SI{0}{K}$ or in the NVT ensemble at $T = \SI{300}{K}$ (\textit{default setup} with edge or mixed 30\degree dislocation only). 
For the \SI{300}{K} simulations, the sample was expanded according to the lattice constant of the Mg matrix at \SI{300}{K} and subsequently equilibrated at \SI{300}{K} for \SI{50}{ps} while keeping the setup stress-free using Nos\'e -Hoover thermostat and barostat~\cite{Nose1984a,Hoover1985, Hoover1986}. The time step for all the MD simulations was set to $\delta t = \SI{1.0}{fs}$.

The lower bound of the CRSS corresponds to the highest applied shear stress for which a static simulation (following the above described procedure and minimization criterion) resulted in a stable equilibrium configuration with a pinned dislocation. 
The upper bound of the CRSS corresponds to the lowest applied shear stress 
for which no stable equilibrium configuration could be found and the dislocation was no longer pinned. 
In the following, the CRSS is provided as the central value between the upper and the lower bound while the error gives the interval between these bounds.



\begin{figure}[t]
	\centering
	\includegraphics[width=\linewidth]{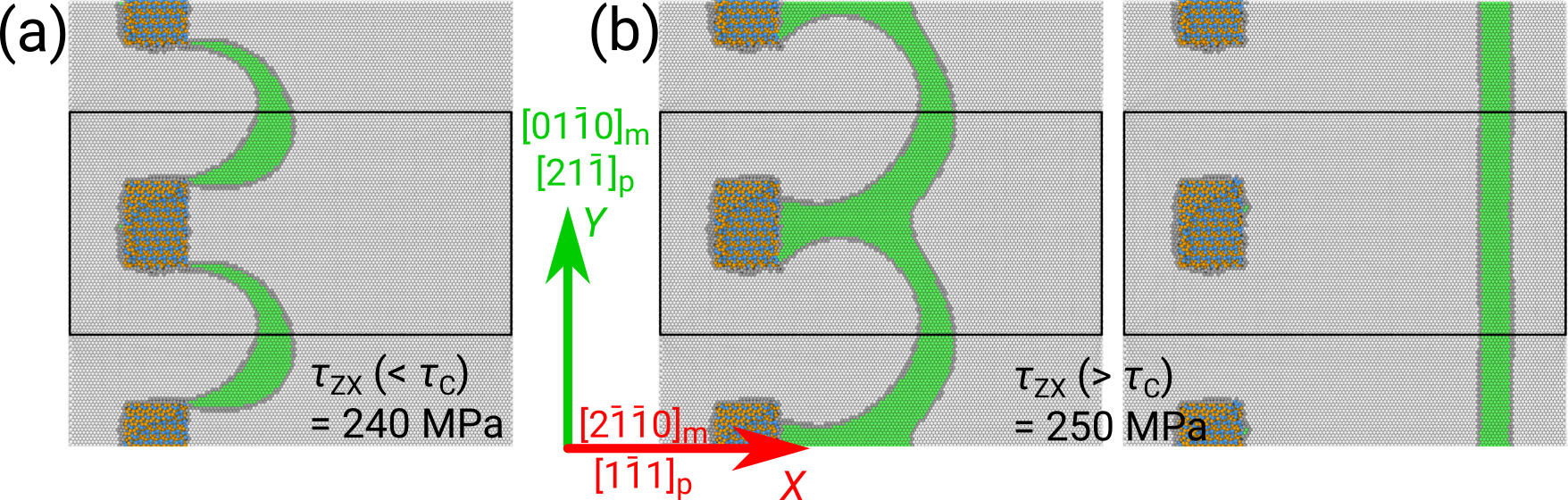}
	\caption{Bracketing of $\tau_\text{c}$ for the \textit{default setup} ($L = \SI{122}{\AA}$, $D = \SI{78}{\AA}$, edge dislocation). 
		(a) relaxed configuration at $\tau_\text{zx} (< \tau_\text{c}) = \SI{240}{\mega\pascal}$.
		(b) snapshots from the pseudo-dynamics of the relaxation process at $\tau_\text{zx} (> \tau_\text{c}) = \SI{250}{\mega\pascal}$.}
	\label{fig:CRSS_Computation}
\end{figure}

\subsection{Visualization and Analysis}
The Open Visualization Tool (OVITO)~\cite{Stukowski2009} was used to visualize and analyze the atomistic configurations. Common neighbor analysis (CNA) ~\cite{Faken1994, Honeycutt1987} was used to identify defects in the Mg matrix. Displacement vector analysis was used to compute the relative displacement between two atomic configurations. Atomic stresses were calculated based on the virial formulation~\cite{Thompson09JCP} and the atomic Voronoi volume~\cite{Zhou2003, Allen1989}. Stresses represented in the current work have been averaged over the nearest neighbors. Atoms within $\approx$ \SI{0.5}{nm} of the IPB were classified  as  "other"  by the CNA algorithm. Although this classification is often associated to amorphous structures, it is important to point out that the present inter-phase boundary (IPB) structure is not amorphous but rather disordered (see section S5 in the supplementary material).


\section{Results}
\label{Results}

\subsection{First dislocation-precipitate interaction}
\label{Results: First dislocation-precipitate interaction}


Upon energy minimization, the dislocation in the \textit{default setup} was attracted to the IPB and was found to be already at the precipitate, even when no stress was applied, see fig. S6 in the supplementary material.
From this relaxed initial structure, the CRSS for a basal edge dislocation to overcome the precipitate in the \textit{default setup} ($L = \SI{122}{\AA}$, $D = \SI{78}{\AA}$) was determined to be $245 \pm \SI{5}{\mega\pascal}$, \textit{i.e.}, at $\tau_{\text{zx}} = \SI{240}{\mega\pascal}$ minimization resulted in a relaxed pinned dislocation, see fig.~\ref{fig:CRSS_Computation}, whereas  at $\tau_{\text{zx}} = \SI{250}{\mega\pascal}$ the dislocation passed the precipitate array. 
The relaxed configuration at $\tau(<\tau_C) = \SI{240}{MPa}$) was furthermore unloaded to 0 stress and a subsequent energy minimization was performed. 
The parts of the dislocations which were absorbed into the IPB remained in the IPB while the rest of the dislocation line assumed a straight shape, see fig. S9 in the supplementary material. 
In some setups (depending on dislocation line-length, dislocation character, precipitate shape), the dislocation was not at the interface upon initial energy minimization with $\tau = \SI{0}{MPa}$. 
These configurations were additionally relaxed with a nominal initial applied stress $\tau = \SI{10}{MPa}$ such that the dislocation was at the matrix-precipitate interface. 
The CRSS, \tauC, for all simulated configurations in tab.~\ref{tab:ParametersSpace} are reported in tab.~\ref{table: Results_CRSS_Values}.

\begin{table}
	\centering
	\caption{ Critical resolved shear stresses \tauC for the studied configurations. Only one parameter is varied at a time, with the others 
	corresponding to the \textit{default setup} (see tab.~\ref{tab:ParametersSpace})}
	\label{table: Results_CRSS_Values}
	\resizebox{\columnwidth}{!}{
	
\small
	\begin{tabular}{llr}
		\toprule
		Varied parameter & Parameter value & \tauC [MPa] \\ 
		\midrule
		Precipitate spacing $L$ (\AA) & 122 & $245 \pm 5$ \\
		 & 222 & $155 \pm 5$ \\
		 & 322 & $115 \pm 5$ \\
		 & 422 &  $95 \pm 5$ \\
		 & 522 &  $75 \pm 5$ \\
		\midrule
		\textit{Precipitate geometry} & & \\
		-- Cuboidal  (width in \AA)
		&  78 & $115 \pm 5$\\
		 & 156 & $135 \pm 5$\\
		& 234 & $145 \pm 5$\\
		-- Spherical (diameter in \AA) 	
		& 20 & $85 \pm 5$\\
		& 50 & $195 \pm 5$\\
		& 78 & $225 \pm 5$\\
		-- Columnar (width in \AA)	
		& 78 & $245 \pm 5$\\
		\midrule
		\textit{Matrix-precipitate interface}& & \\
		-- orientation relationship & Burgers OR & $245 \pm 5$\\
		& \hkl[1-11]$_{p}$ $\parallel$ \hkl[2-1-10]$_{m}$, \hkl(21-1)$_{p}$ $\parallel $\hkl(0001)$_{m}$ & $215 \pm 5$ \\
		& \hkl[011]$_{p}$ $\parallel$ \hkl[2-1-10]$_{m}$, \hkl(-11-1)$_{p}$ $\parallel $\hkl(0001)$_{m}$ & $215 \pm 5$ \\
		-- slip plane alignment 
		& \hkl(0001)$_m$/\hkl(011)$_p$ & $245 \pm 5$\\
		& \hkl(0001)$_m$/\hkl(033)$_p$ & $245 \pm 5$\\
		-- termination (type) 
		& I1 or I3 & $245 \pm 5$ \\
		& I2 or I4 & $235 \pm 5$\\
		\midrule
		\textit{Dislocation character} \\
		-- edge   
		& \hkl(0001)$_m$/\hkl(011)$_p$ & $245 \pm 5$ \\
		& \hkl(0001)$_m$/\hkl(033)$_p$ & $245 \pm 5$\\
		-- 30\degree 
		& \hkl(0001)$_m$/\hkl(011)$_p$ & $275 \pm 5$\\
		& \hkl(0001)$_m$/\hkl(033)$_p$ & $275 \pm 5$\\
		\midrule
		Obstacle type
		& void & $195 \pm 5$ \\ 
		& precipitate & $245 \pm 5$ \\
		& frozen atoms & $285 \pm 5$\\
		\bottomrule
	\end{tabular}}
\end{table}

\begin{figure*}[h]
	\centering
	\includegraphics[width=\linewidth]{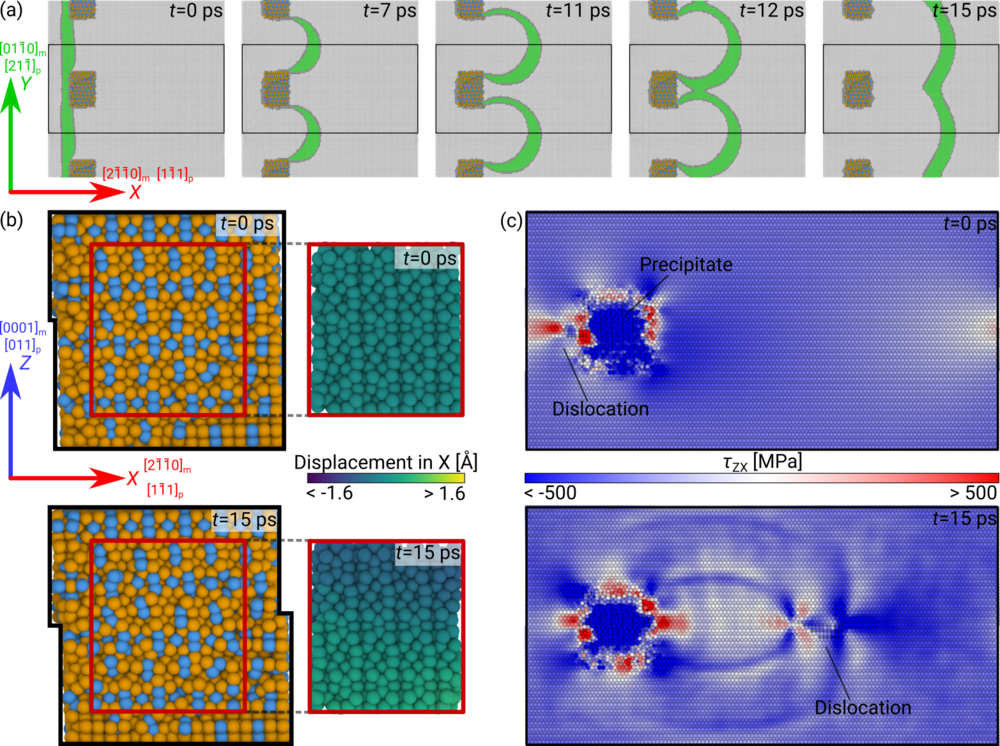}
	\caption{(a) Snapshots of an infinite straight basal edge dislocation interacting with a periodic array of \MgAl{} precipitates, as simulated by molecular dynamics (NVE at $T_0 = \SI{0}{\kelvin}$, $\tau_{\text{zx}} = \SI{250}{MPa}$). See fig.~\ref{fig:SimulationSetup} for the color coding.
		(b) \MgAl{} precipitate before ($t=\SI{0}{\pico\second}$) and after ($t=\SI{15}{\pico\second}$) interaction with an edge dislocation. Please note the step at the matrix-precipitate interface. The insets show the  precipitate interior with the color corresponding to the displacement of the atoms along the direction of dislocation motion with respect to the configuration at $t=0$ ps. 
		(c) stress field $\tau_{\text{zx}}$ before ($t=\SI{0}{\pico\second}$) and after ($t=\SI{15}{\pico\second}$) interaction with an edge dislocation.}
	\label{fig:Mechanism}
\end{figure*}

Snapshots of the interaction of an initially straight edge dislocation at $\tau_{\text{zx}}= \SI{250}{\mega\pascal}$ with the precipitates in the \textit{default setup} are shown in fig.~\ref{fig:Mechanism}(a) (NVE, $T_0=0 \text{K}$). 
At $t=\SI{0}{\pico\second}$, the dislocation started at the leftmost matrix-precipitate interface. 
The dislocation moved further to the right and at $t=\SI{7}{\pico\second}$  bowed out between the precipitates. 
At $t=\SI{11}{\pico\second}$, the dislocation reached a maximum bow-out state. 
The dislocation arms on both sides of the precipitate  attracted  each other and annihilated at $t=\SI{12}{\pico\second}$, after which the dislocation depinned from the precipitate ($t=\SI{15}{\pico\second}$). 
Using the common neighbor analysis, no dislocation loops were visible in the matrix around the precipitate, both in the static and in the dynamic simulations at different temperatures.

\begin{figure*}
	\centering
	\includegraphics[width=\linewidth]{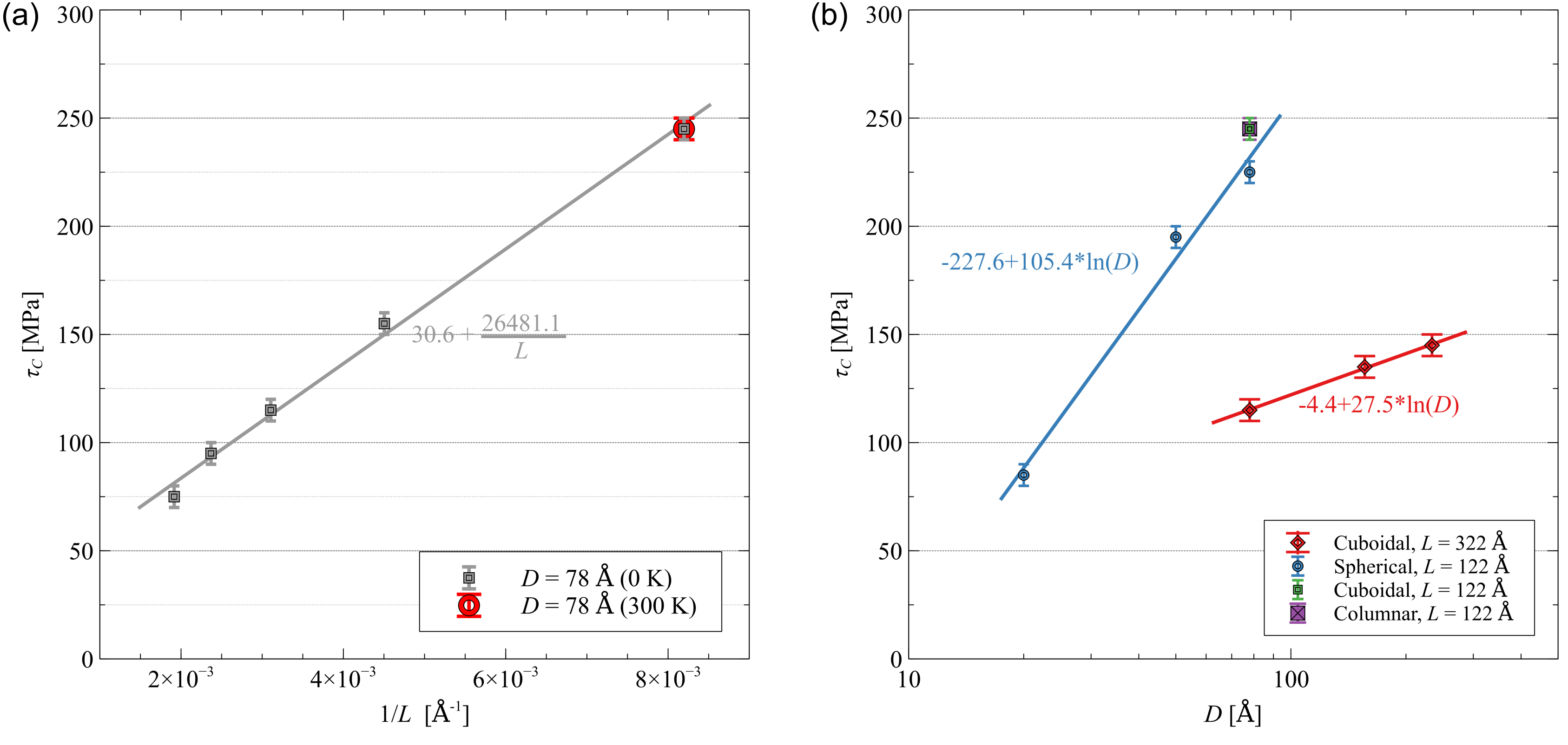}
	\caption{Dependence of the critical resolved shear stress (CRSS) on (a) inter-precipitate spacing $L$ and (b) precipitate size. }
	
	\label{fig:Effect of L and D}
\end{figure*}

Figure~\ref{fig:Effect of L and D}(a) shows the CRSS for different inter-precipitate spacings, $L$. 
Increasing the inter-precipitate distance resulted in a smaller value of \tauC. 

To study the effect of precipitate shape, spherical, rectangle-columnar (spanning the height of the simulation box) and cuboidal (as used for the \textit{default setup}) geometries were used in the simulations.
Note that changing the precipitate shape also implies that the matrix-precipitate interfaces were crystallographically different. 
The CRSS for a spherical precipitate was lower than for cuboidal or columnar precipitates (see fig.~\ref{fig:Effect of L and D}(b), $L = \SI{122}{\AA}$). 
The mechanism by which the dislocation passes the precipitates was the same for all precipitates, except for the spherical one with a diameter of 20~\AA, which was cut by the incoming dislocation. 

Figure~\ref{fig:Effect of L and D}(b) also shows the effect of varying precipitate size (width for cuboidal,  diameter for spherical precipitates)
 on \tauC . 
Instead of the default value of $L = \SI{122}{\AA}$ a larger value of $L = \SI{322}{\AA}$ was constantly used for a cuboidal precipitate.
As can be seen in fig.~\ref{fig:Effect of L and D}(b) the CRSS showed a  logarithmical dependence with the precipitate size.

To study the influence of the matrix-precipitate interface structure on the dislocation-precipitate interaction, the interface was modified in three ways: 
(1) by rotating the precipitate and thus, changing the orientation relationship between the matrix and the precipitate, 
(see tab.~\ref{table: Results_CRSS_Values})
(2) by changing the slip plane of the matrix dislocation so that it aligns with the  \hkl{330} plane of the \MgAl{} precipitate instead of the \hkl{110} plane, and
(3) by changing the termination plane of the precipitate at the matrix-precipitate interface (see supplementary material section S2).
 Changing the OR between the matrix and precipitate may change the effective inter-precipitate distance and the precipitate width, leading to a lower CRSS. 
 Changing the relative alignment 
 of the slip plane of the matrix dislocation  with respect to the precipitate did not influence the CRSS. 
 Changing the termination plane of the precipitate at the matrix-precipitate resulted in only slight variations in \tauC as can be seen in tab.~\ref{table: Results_CRSS_Values}. 

\begin{figure*}
	\centering
	\includegraphics[width=\linewidth]{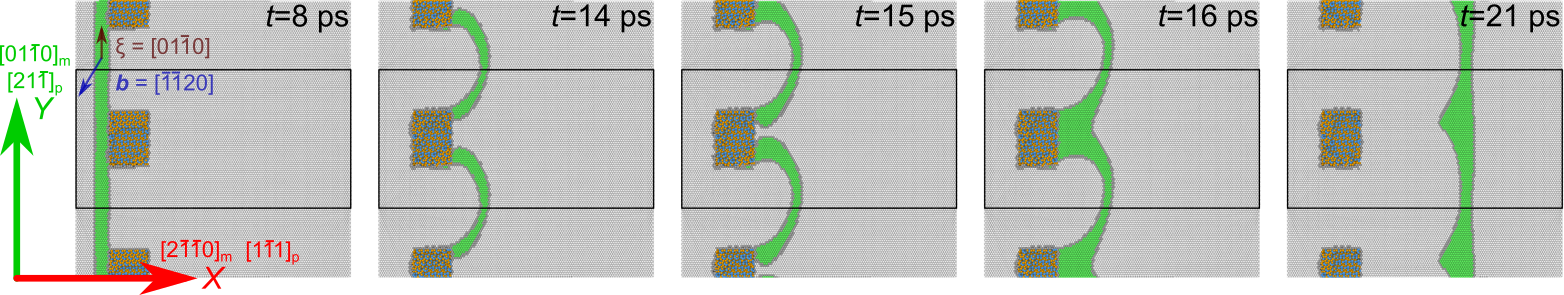}
	\caption{Snapshots of an infinite straight basal 30\degree\ dislocation interacting with a periodic array of \MgAl{} precipitates, as simulated by molecular dynamics (NVE at $T_0 = \SI{0}{\kelvin}$, $\tau_{\text{applied}} = \SI{280}{MPa}$). Color coding, see Fig.~\ref{fig:SimulationSetup}.}
	\label{fig:Effect dislocation character}
\end{figure*}

To study the influence of the character of the dislocation on its interaction with the precipitate, a 30\degree{} dislocation was introduced 
 in the matrix instead of the default edge dislocation.
Forces equivalent to a shear stress greater than \tauC were applied to the system along the direction of the Burgers vector. 
Figure~\ref{fig:Effect dislocation character} shows 
the interaction of a 30\degree{} dislocation with \MgAl{} precipitates in an NVE simulation at $\SI{0}{\kelvin}$. 
The shown stages of dislocation-precipitate interactions are the same 
as in fig. \ref{fig:Mechanism}(a), however, as the dislocation did not approach the precipitate during energy minimization, the times in both figures are not identical.
At $t=\SI{8}{\pico\second}$, the dislocation arrived at the matrix-precipitate interface, and continued to move along the sides of the precipitates (see $t=\SI{14}{\pico\second}$). 
At $t=\SI{15}{\pico\second}$, the dislocation reached its maximum bow out state. 
Before overcoming the precipitate, the dislocation arms interacted and annihilated each other ($t=\SI{16}{\pico\second}$) before the dislocation finally detached ($t=\SI{21}{\pico\second}$). 
In contrast to the edge dislocation, fig.~\ref{fig:Mechanism}(a), 
the maximum bow-out configuration of the 30\degree{} dislocation was asymmetric, but the overall mechanism of interaction was similar to the edge dislocation case. 
The CRSS  for the 30\degree{} dislocation  was about 10\% larger than that for an edge dislocation (see tab.~\ref{table: Results_CRSS_Values}).
Similar to the edge dislocation case, changing the alignment of the slip plane did not affect the CRSS.

We performed additional simulations of selected configurations at \SI{300}{\kelvin} and found no change of interaction mechanism for either edge or  30\degree{} dislocations (see supplementary material section S6), and the CRSS of the  \textit{default setup}  was - within the bracketing interval -  identical to the CRSS at \SI{0}{K} (see figure \ref{fig:Effect of L and D}).



\begin{figure}[!t]
	\includegraphics[width=\linewidth]{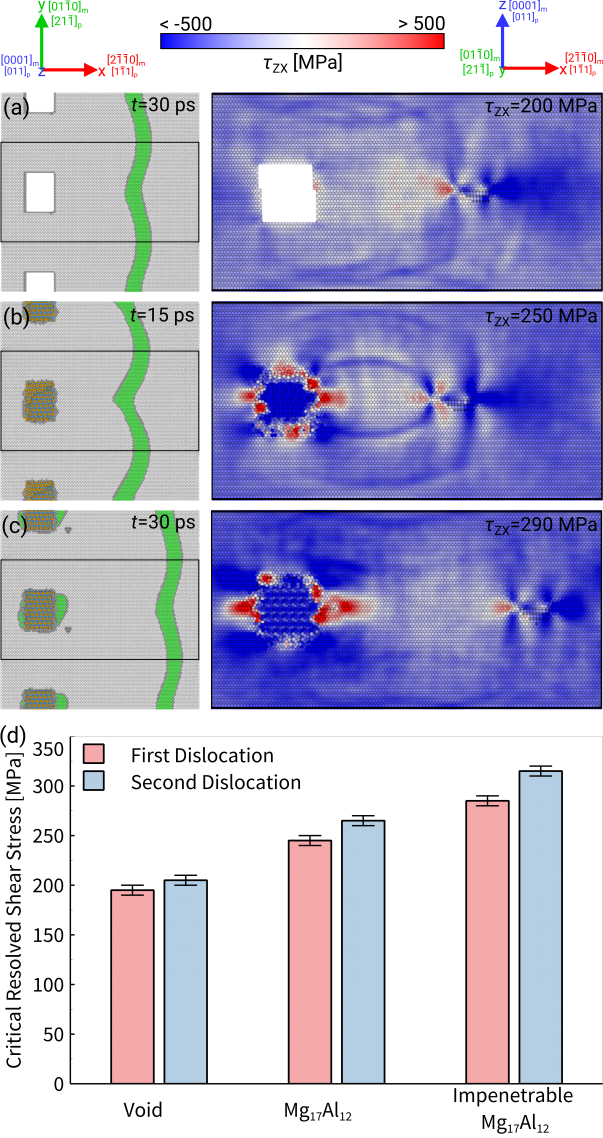}
	\caption{Top view of the setup (left) and side view of the stress field $\tau_{\text{zx}}$ (right) after a dislocation has overcome an array of (a) voids, (b) \MgAl{} precipitates, and (c) impenetrable \MgAl{} precipitates the first time studied using NVE at $T_0 = \SI{0}{\kelvin}$ ($L = \SI{122}{\AA}$, $D = \SI{78}{\AA}$). (d) Variation in critical resolved shear stress with modification of obstacle type. }
	\label{fig:Modification of obstacle}
\end{figure}

Two extreme cases of obstacles were studied in addition to the 
default \MgAl{} precipitate: an impenetrable \MgAl{} precipitate achieved by fixing the corresponding atomic positions, and a void, both of the same shape and size as the default precipitate.
Figure~\ref{fig:Modification of obstacle} shows snapshots from the corresponding MD simulations (NVE, $T_0=\SI{0}{\kelvin}$) when the edge dislocation has passed the obstacle.
In the case of the void, fig.~\ref{fig:Modification of obstacle}(a), 
the dislocation was attracted to the void. The part of the dislocation touching
the void disappeared at the free surface leaving a step, while the remaining dislocation moved along the void, bowed out with the side arms annihilating and creating a step at the back surface of the void in the process.
With the fixed \MgAl{}  precipitate (fig.~\ref{fig:Modification of obstacle}(c)), the dislocation overcame the precipitate by Orowan looping, similar to the already addressed case of the default \MgAl{} precipitate fig.~\ref{fig:Modification of obstacle}(b)
In this case, a dislocation loop in the matrix was clearly detected by the CNA algorithm.
The corresponding stress states after obstacle passage are also shown on the right of fig.~\ref{fig:Modification of obstacle}.

The CRSS to pass the void was 20\% lower than for the \MgAl{} precipitate of identical size. The impenetrable \MgAl{} precipitate, on the other hand, required an about 15\% higher resolved shear stress than the \MgAl{} precipitate with mobile atoms, see 
fig.~\ref{fig:Modification of obstacle}(d).

\subsection{Second dislocation-precipitate interaction} 
\label{Results: Second dislocation-precipitate interaction}

After having overcome the obstacle, the dislocation can pass through the PBC and interact a second time with the same obstacle. Quasi-static simulations were performed iteratively by increasing the applied shear stress by \SI{10}{\mega\pascal} until the dislocation overcame the obstacle during energy minimization, using the previous relaxed configuration. The inertial effects from the pseudo-dynamics of FIRE~\cite{Bitzek2005,Bitzek2006} were avoided as the starting configuration was relaxed with the dislocation next to the precipitate.
The CRSS for the second interaction are shown in fig.~\ref{fig:Modification of obstacle}(d).
While in case of the void, the CRSS for the second edge dislocation passing was within the error margin of the CRSS for the first passage, the CRSS for the 
second dislocation-precipitate interaction in the case of the  \MgAl{} precipitate was about 8\% higher than for the first dislocation passing. 
The increase in resolved shear stress necessary to pass the impenetrable \MgAl{} precipitate a second time was 10\% higher --- slightly more pronounced than for the \MgAl{} precipitate with mobile atoms.

\begin{figure*}
	\centering
	\includegraphics[width=\linewidth]{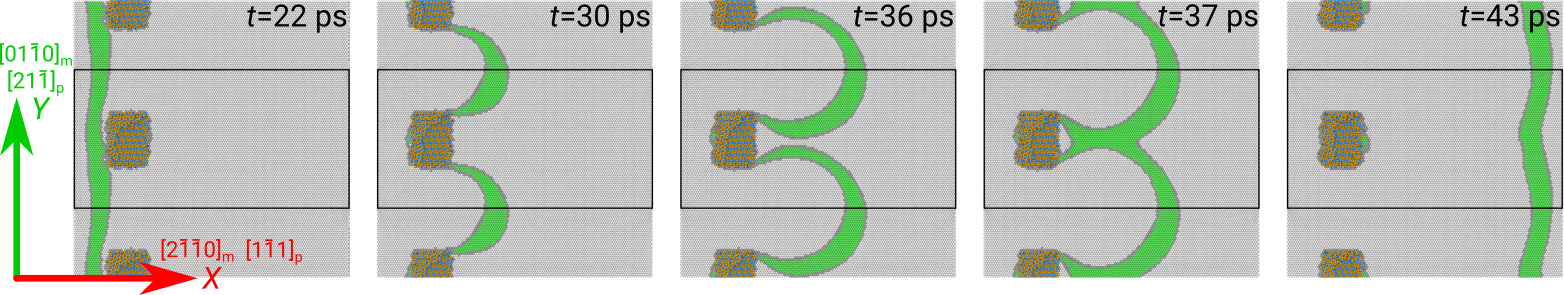}
	\caption{Snapshots of a second infinite straight basal edge dislocation interacting with a periodic array of \MgAl{} precipitates for the \textit{default setup}(NVE at $T_0 = \SI{0}{\kelvin}$, $\tau_{\text{zx}} = \SI{270}{MPa}$). See fig.~\ref{fig:SimulationSetup} for the color coding.}
	\label{fig:Effect of Second Dislocation Pass}
\end{figure*}

Snapshots from the MD simulation of the second dislocation interaction with the precipitate (NVE, $T_0 = \SI{0}{\kelvin}$, $\tau= \SI{270}{\mega\pascal}$) 
are shown in fig.~\ref{fig:Effect of Second Dislocation Pass}. Here, the starting configuration was relaxed at $\tau = \SI{0}{\mega\pascal}$ and the dislocation was next to the precipitate. The applied shear stress was larger than the CRSS; therefore the dislocation overcame the precipitate and re-interacted with it. In this case, the inertial effects during the second dislocation interaction with the precipitate cannot be neglected.
As can be seen from fig.~\ref{fig:Effect of Second Dislocation Pass}, the mechanism of interaction is the same as for the first passage, fig.~\ref{fig:Mechanism}. 
Small dislocation segments remaining in front and in the back of the precipitate can be identified, see fig.~\ref{fig:Effect of Second Dislocation Pass}, which were subsequently absorbed.

The relaxed interface structure before and after absorption of the first and second dislocation are shown in fig. S5 in the supplementary material.
The step at the interface where the edge dislocation was absorbed is clearly visible and increasing with the absorbed dislocation contents, while remaining disordered. 
Any visible changes in the IPB structure were confined to the region where the dislocation was absorbed (see fig. S5 in the supplementary material).

\section{Discussion}
\label{Discussion}

The absence of any visible dislocation loop around a precipitate after 
dislocation passage and the presence of a step at the precipitate-matrix interface is commonly taken as an indication that the dislocation sheared the precipitate. 
Indeed, Moitra and LLorca \cite{Moitra2017} used these criteria to determine that the \SI{2}{nm} to \SI{10}{nm} \MgAl{} disc-shaped precipitates in Burgers orientation relationship, which they simulated, 
were sheared by a basal edge dislocation inserted in the Mg Matrix. 
Liao \etal \cite{Liao2014}, on the other hand, found that for a \SI{5}{nm} 
\MgAl{} precipitate in the same orientation relationship, the edge dislocation only caused \emph{elastic} deformation of the precipitate.
These studies were, however, performed with different potentials than in
the present simulations, that have certain deficiencies when coming to model the Mg-\MgAl{} system, see section S1 of the supplementary material.

Figure~\ref{fig:Mechanism}(b) shows clear steps at the precipitate after dislocation passage.
However, when only considering the precipitate interior, the displacement field shows no plastic deformation after the dislocation passed the precipitate, see the insets in fig.~\ref{fig:Mechanism}(b).
Additionally, analysis of the stress field around the precipitate, fig.~\ref{fig:Mechanism}(c), reveals at $t=$\SI{15}{ps}  a remaining stress signature in front and in the back of the precipitate similar to the one of the dislocation in front of the precipitate ($t=$\SI{0}{ps}).
These are clear indications that instead of leaving an Orowan loop in the matrix around the precipitate, the dislocation was absorbed in the Mg/\MgAl{} interface.
The absence of Orowan loops and the presence of steps at the precipitate-matrix interface can, therefore, not be used to infer that precipitates were sheared.

It is well known that interphase boundaries (IPB), similar to grain boundaries, can act as sinks for dislocations \cite{Sutton1995, Wolf2005, Dao2007, Wang2011, Bitzek2009} 
We observed that dislocation absorption took place in all of our simulations with \MgAl{} precipitates, except the smallest one, independent of the details on how exactly the matrix-precipitate interfaces were constructed (different terminations, precipitate shapes and orientations, see tab.~\ref{tab:ParametersSpace}), dislocation character or the simulation temperature. 
This fact points to dislocation absorption in the IPB as generic feature for basal dislocations in Mg interacting with \MgAl{} precipitates.
Liao \etal \cite{Liao2014}, however, found for a 3 nm \MgAl{} precipitate a different interaction mechanism, which they interpreted as cross-slip of their basal edge dislocation. 
Their observed resulting dislocation configuration after the edge dislocation passed the precipitate
with two super-jogs could also be explained by the absorption by the dislocation of vacancies from an excess free volume at the interface. 
In this case, the edge dislocation would have been effectively \emph{climbing} over the precipitate.
The way the IPBs are created in the atomistic simulations can therefore have important consequences on the dislocation precipitate interaction mechanisms. 

High-resolution transmission electron microscopy (HR-TEM) analysis of Mg-\MgAl{} interfaces shows local distortions at the interphase boundary \cite{Wang2003, Hutchinson2005}. Although, these images may not directly resolve the local structure, the loss of contrast from distinct atomic columns indicates that the interface is not perfectly coherent despite the orientation relationship being maintained between the two phases. This observation agrees with the disordered IPB structure in our simulations (see supplementary material section S5) 
An IPB with excess free volume (see also fig. S4 in the supplementary material) facilitates the absorption of dislocations. 
The IPB thickness of approx. \SI{0.5}{nm} surrounding the particle in our simulations
also explains why the smallest \SI{2}{nm} spherical precipitate was sheared in 
our simulations, as nearly most of its atoms were part of the IPB.

In order for dislocations to be absorbed in the Mg/\MgAl{} IPB, the image force on the dislocation caused by the different elastic constants of the precipitate and the matrix \cite{Hirth1982} has to be either negligible or attractive. 
A repulsive image force as in the case of the fixed \MgAl{} would cause a certain stand-off distance of 
the dislocation, resulting in a visible Orowan loop as can be seen in fig.~\ref{fig:Modification of obstacle}(c). Using the shear modulus in slip plane and slip direction with the values for the used potential shows that $\mu'($Mg$)= \SI{17.2}{GPa} \approx \mu'($\MgAl{}$)= \SI{17.6}{GPa}$ (see supplementary material section S4). Comparing the elastic constants from MEAM potential with DFT and experiments for both phases, no differences are expected (supplementary material section S1).

From the fact that the precipitate in the \textit{default setup} is not sheared by a second dislocation, one can deduce a lower bound for the CRSS to shear the \MgAl{} by assuming a pile-up of $n=2$ dislocations at an applied shear stress of $\tau= \SI{250}{MPa}$, fig.~\ref{fig:Modification of obstacle}(d), to $\tau_\text{c,min}=n\tau=\SI{500}{MPa}$. 
This is consistent with the higher stress required to move dislocations in the complex intermetallic phase \MgAl{} compared with the pure metallic Mg matrix. 
At room temperature, the hardness of \MgAl{} has been measured as \SI{2}{GPa} and \SI{3.5}{GPa} in micro- and nano-hardness measurements, respectively ~\cite{Mathur2016, Fukuchi1975}. 
In contrast, the micro-hardness of pure magnesium has been measured to \SI{0.34}{GPa}~\cite{Edalati2011}. 
The much higher hardness of the intermetallic phase is also evidenced by its brittle behavior at room temperature, where fracture becomes dominant in macroscopic mechanical testing ~\cite{Song2009, Ragani2011, Maghsoudi2015, Hagihara2018}.
This is consistent with the tetrahedral packing leading to corrugated crystallographic planes in \MgAl{} and the correspondingly large perfect Burgers vector expected on these planes which lead to limited thermal activation of dislocation glide, as found in hardness measurements between room temperature and \SI{0.54}{T_m} ~\cite{Mathur2016}. 
Although Xiao \etal ~\cite{Xiao2013} suggested a specific \hkl(110) slip plane
and microcompression experiments have confirmed slip traces consistent with this type of plane ~\cite{Kolb2013}, the actual Burgers vector, including that of any partial dislocations and the exact slip planes of \MgAl{} have, however, not yet been unambiguously identified.\\

The critical resolved shear stress  in dislocation - obstacle interactions can be significantly reduced by so called 'inertial overshooting',  particular at low temperatures and high dislocation velocities~\cite{Bitzek2006,Bitzek2005}. 
As FIRE uses  0K pseudo-dynamics ~\cite{Bitzek2006}, similar inertial effects can in principle occur during minimization if the dislocations are initially far away from the obstacles.
In the present study, however, no such inertial effects are expected as the dislocations were close to the precipitates when the CRSS was determined.

The critical resolved shear stress $\tau_c$ to unpin a dislocation from a periodic array of 
strong obstacles like voids or impenetrable particles has been shown to be dominated by the stress necessary to pull out the dislocation segments at the obstacle into a parallel dipole \cite{Bacon1973}.
Bacon, Kocks and Scattergood \citep{Bacon1973} and Scattergood and Bacon~\cite{Scattergood1982} 
developed the following expression for  $\tau_c$ by fitting their results obtained by computer simulations based on elasticity theory, which explicitly modeled the self-stress of a flexible dislocation:
\begin{equation}
	\label{Equation by Scattergood and Bacon}
	\tau_{c} = \frac{\mu'b}{2\pi A L} \Big(\ln(\bar{D}) + \Delta \Big) .
\end{equation}
In this expression, which will henceforth be referred to as the BKS model, $\mu'$ is the shear modulus in slip plane and slip direction, $A$ equals 1 for an initially pure edge dislocation and $(1-\nu)$ for a pure screw dislocation, where $\nu$ is Poisson's ratio, $b$ is the magnitude of the Burgers vector and $\bar{D}$ is the harmonic mean of the inter-obstacle spacing $L$ and the obstacle width $D$: $\bar{D}=(\nicefrac{b}{L} + \nicefrac{b}{D})^{-1}$. 
The empirical constant $\Delta$ models the energetic costs associated with the specific dislocation-obstacle interaction.
In the case of a void, it describes the resisting force caused by creation of a surface step \cite{Scattergood1982}:
\begin{equation}
	\Delta = \frac{\delta\gamma}{\frac{\mu'b}{4 \pi } \ln\big(\nicefrac{R}{r_0}\big)} .
\end{equation}
Here, $\delta\gamma$ is the energy of the newly created interface (in case of a void, $\delta\gamma$ equals $\gamma_\text{surf}$, the energy of a free surface) and $R, r_0$ are the outer and inner cut-off radii in the calculation of dislocation energy (the factor $ln(\nicefrac{R}{r_0})$ is usually taken to be unity).

\begin{figure}
	\includegraphics[width=\linewidth]{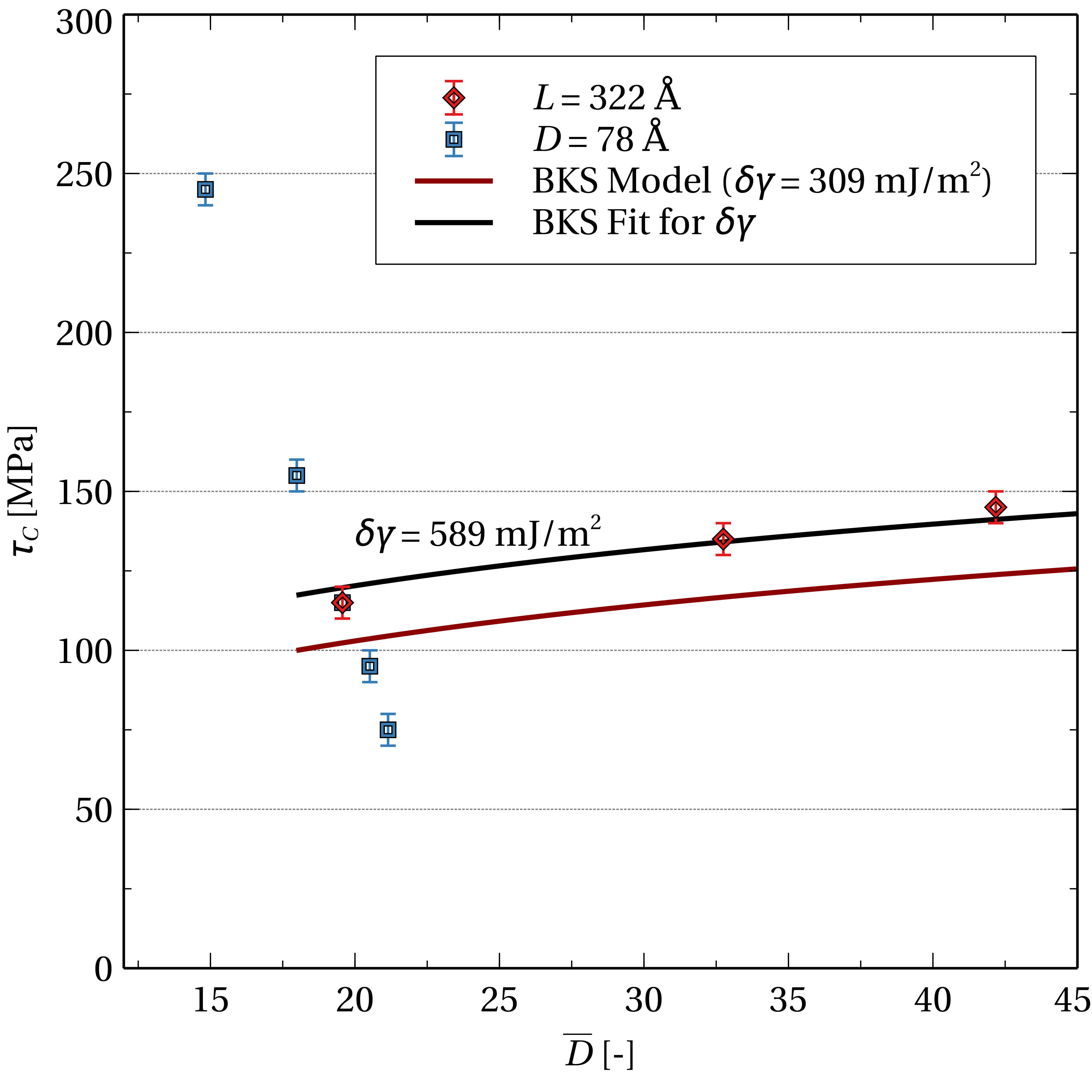}
	\caption{Critical resolved shear stress $\tau_c$ for an edge dislocation to pass \MgAl{} precipitates with a constant spacing $L$ and varying diameter $D$, as well as constant $D$ and varying $L$ plotted against the harmonic mean, $\bar{D}$. 
		Atomistic simulations (data points), BKS model using $\delta\gamma=\gamma_{\text{interface}}$ 
		(red line) and with fitted $\delta\gamma$ (black line). }
	\label{fig:Model by Scattergood and Bacon}
\end{figure}

Figure~\ref{fig:Model by Scattergood and Bacon} shows the values of $\tau_c$ determined from the simulations of an edge dislocation interacting with \MgAl{} precipitates of various diameters and spacing 
as function of $\bar{D}$.
Using the BKS model with $\mu' = \SI{17.2}{\giga\pascal}$, $b = \SI{3.204}{\AA}$ and 
an estimate for the interface step energy $\delta\gamma=\gamma_\text{interface} = \SI{309}{\milli\joule\per\metre\squared}$ calculated for a \hkl(011)$_{p}$ $\parallel $\hkl(0001)$_{m}$ 
 interface (following the same methodology as in \cite{Wang2018}) results in too low values for $\tau_c$ compared to the simulations. 
Fitting the BKS model to the data results in $\delta\gamma=\SI{589}{\milli\joule\per\metre\squared}$,
i.e., the energy of the absorbed dislocation in the IPB is much larger than the energy created by interfacial steps of width $b$. 

The energy of the absorbed dislocation in the IPB, modeled by $\Delta$ in eq. ~\ref{Equation by Scattergood and Bacon}, can be influenced by the temperature. 
In the simulations, no  effect of temperature was observed on the CRSS at 300K (see figure \ref{fig:Effect of L and D}) for $L = \SI{122}{\AA}, D = \SI{78}{\AA}, \bar{D} = 14.83$.
On experimental time scales the dislocation core  might spread and dissolve within the IPB, 
which could lead to some temperature dependence of the CRSS. 
The main contribution to the obstacle strength is however the elastic energy due to the dislocation dipole configuration, which should not be strongly dependent on temperature.


Interestingly, the BKS model predicts a larger $\tau_c$ for voids than for the Orowan process 
($\Delta(\text{Orowan})= 0.77, \Delta(\text{void})=1.52$ \cite{Bacon1973,Scattergood1982}.
Our simulations with $D=\SI{78}{\AA}$ and $L=\SI{122}{\AA}$ ($\bar{D}=14.83$), however, show that the void has $\approx$ 0.8 times smaller $\tau_c$ than an \MgAl{} precipitate with identical $D$ and $L$, see fig.~\ref{fig:Modification of obstacle}(d).
This points to the importance of the specific details of the precipitate and the IPB, which were not 
accounted for in the derivation of the BKS model~\cite{Bacon1973} and can become particularly important for small $\bar{D}$. 

This becomes particularly evident when considering the fixed \MgAl{} precipitate, which has a $\approx$ 1.2 times higher $\tau_c$ than the \MgAl{} precipitate. 
Here, the infinitely stiff obstacle leads to high image forces and subsequently a higher energy of the Orowan loop, which are not accounted for in the original BKS model with non-interacting obstacles of identical elastic properties as the matrix \cite{Bacon1973}.
Their model of a flexible dislocation, however, captures the importance of the Burgers vector orientation relative to the obstacle row and predicts a higher $\tau_c$ for mixed dislocations as compared to a pure edge dislocation, in agreement with our results for a 30\degree\  dislocation, see tab.~\ref{table: Results_CRSS_Values}.

\begin{figure}[t]
	\includegraphics[width=\linewidth]{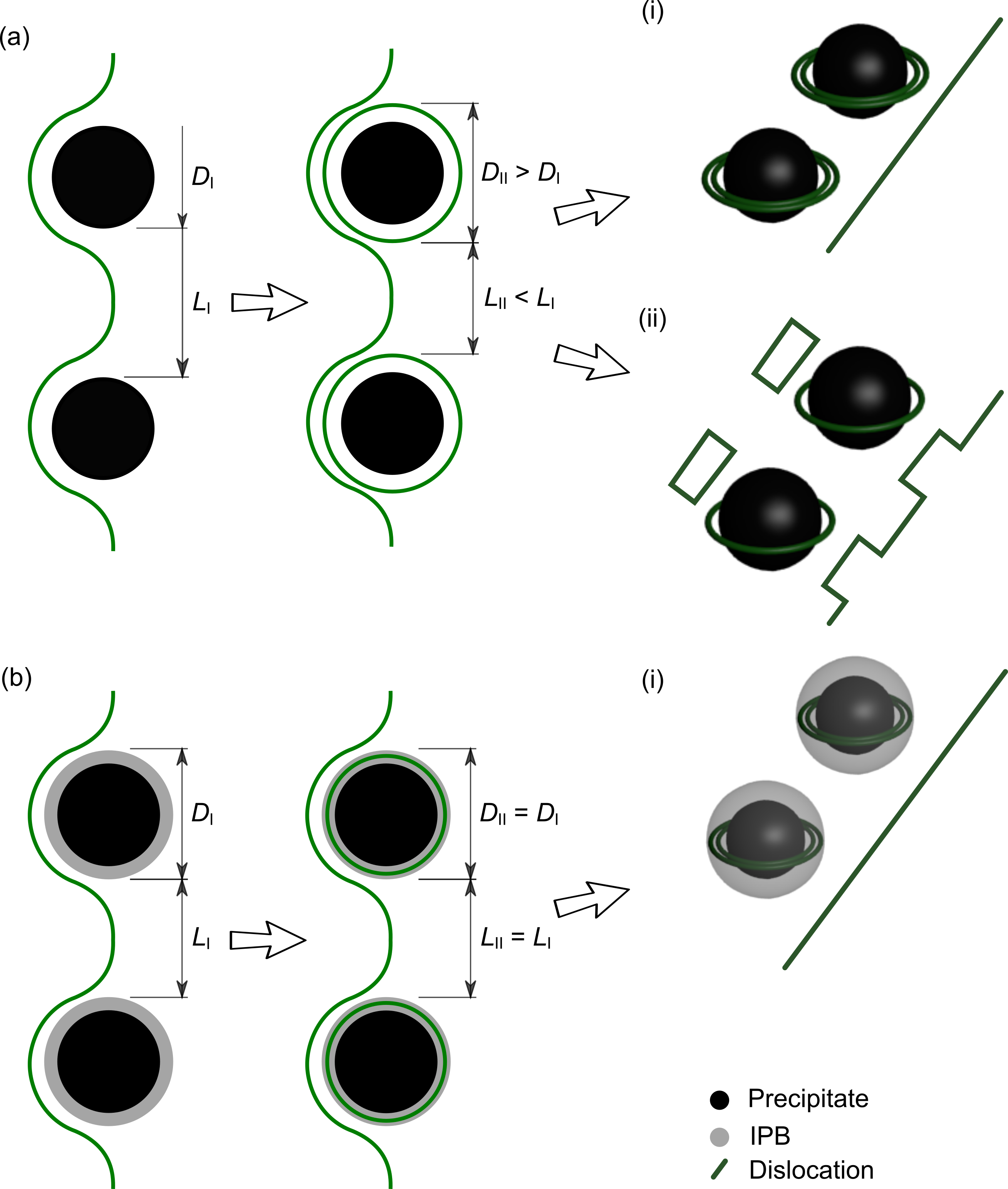}
	\caption{Schematics of the mechanisms of an edge dislocation interacting with impenetrable obstacles: (a) Orowan mechanism (after \cite{Fisher1953}), leading to one or multiple loops in the matrix around the precipitate (i), combined cross-slip-Orowan mechanism according to Hirsch \cite{Hirsch1957, Humphreys1970} (ii);  (b) possible interaction mechanisms when the dislocation is absorbed in the IPB. (Effective) inter-precipitate spacing $L_\text{I}$ ($L_\text{II}$) experienced during the first (second) dislocation-precipitate interaction and (effective) precipitate diameter $D_\text{I}$ ($D_\text{II}$) experienced during the first (second) dislocation-precipitate interaction.}
	\label{fig: Schematic Precipitation Hardening}
\end{figure}

The BKS model does not consider the interaction of multiple dislocations with the same obstacle,
i.e., the change of $\tau_c$ upon repeated bowing or cutting, which would reflect as hardening or softening during ongoing plastic deformation.
In agreement with the simulation results on voids, fig.~\ref{fig:Modification of obstacle}(d), no significant change in $\tau_c$ is expected for voids that are not completely sheared apart,
as the increase in surface energy $\gamma_\text{surf}$ is not expected to change much with the step size. 
For the Orowan process, however, the Orowan loops left around the obstacle by previously interacting dislocations affect subsequent dislocations.
I.e., the work hardening rate for metals and alloys containing unshearable particles is different from the same material without particles or with shearable precipitates \cite{Kelly1971, Brown1971,Queyreau2010}. 
The Orowan loops contribute to short- and long-range interactions with other dislocations \cite{Kelly1971}. 
The short-range interaction reduces the effective inter-particle distance and has a significant strengthening effect. 
This so-called source-shortening effect \cite{Brown1971} was recently studied by dislocation dynamics simulations, where Queyreau \etal \cite{Queyreau2010} could show that this effect of the accumulated Orowan loops can indeed be modeled by the classical approach of assuming an increased effective particle diameter $D_\text{II}>D_\text{I}$ and correspondingly a smaller $L_\text{II}$ in Eq.~\ref{Equation by Scattergood and Bacon} \cite{Courtney2000}, as illustrated in  fig.~\ref{fig: Schematic Precipitation Hardening}(a).

The dislocation absorbed in the IPB clearly affects the second dislocation and leads to an increase in $\tau_c$ for the following dislocation, see figs.\ref{fig:Modification of obstacle}(d) 
and \ref{fig:Effect of Second Dislocation Pass}.
However, due to the absorption in the IPB, the effective particle diameter will not be the same as in the case of an Orowan loop in the matrix, even for otherwise identical particles. 
Furthermore, the line energy and stress field of dislocations can change upon absorption \cite{Gao2002, Bitzek2009, Wang2011, Huang2018}, see also the stress fields in figs.~\ref{fig:Modification of obstacle}(b) vs. \ref{fig:Modification of obstacle}(c).
The contribution to work hardening by precipitates with absorbing interfaces is therefore expected to be different than for particles with interfaces that cannot absorb dislocations, even if other particle properties remain identical. 
Changes of dislocation properties in interfaces were first considered in the case of diffusive processes \cite{Srolovitz1984, Blum1985, Gao2002}, where it helped to explain the creep behavior of oxide-dispersion strengthened (ODS) superalloys.
The recent observations of dislocation-precipitate interactions by Huang \etal \cite{Huang2018}, however, suggest that diffusion-less interfacial dislocation relaxation processes can lower the line energy of dislocations absorbed in an IPB.  

A further consequence of dislocations being absorbed in particle interfaces is that in contrast to Orowan loops in the matrix, they cannot take part in cross-slip processes. 
The combined cross-slip, Orowan mechanisms as suggested by Hirsch \cite{Hirsch1957, Humphreys1970}, see also fig.~\ref{fig: Schematic Precipitation Hardening}(a)(ii), can therefore not take place. Figure~\ref{fig: Schematic Precipitation Hardening} summarizes the different mechanisms possible in the case of Orowan looping in the matrix and dislocation absorption into the IPB.

In the present simulations, only dislocations in Mg interacting with \MgAl precipitates were shown to be absorbed into the IPB.
However, the recent in-situ TEM indentation experiments on Mg-Nd alloys by Huang \etal \cite{Huang2018}
showed dislocation absorption at the interface of $\beta_1$ precipitates of the DO$_3$ structure.
Dislocation absorption into precipitate IPBs might therefore be more common for incoherent particles.
The implications, like changes in dislocation energy and stress field, and unavailability of dislocations for cross-slip processes, are, however, to our knowledge not yet included in mesoscale models of particle hardening.

\section{Summary}
\label{Conclusions}

Molecular statics and dynamics simulations were performed on basal edge and 30\degree  dislocations in Mg interacting with an array of \MgAl\ precipitates to determine the critical stress for obstacle passing and study the mechanisms of dislocation-precipitate interactions. 
Independent of temperature (up to $\SI{300}{K}$), precipitate diameter ($\SI{20}{\AA}<D<\SI{235}{\AA}$), precipitate shape, precipitate orientation, precipitate spacing, dislocation type, the relative alignment of slip planes in matrix and precipitate, and the preparation of the matrix-precipitate interface, the dislocations leave no visible Orowan loop in the matrix around the precipitate. 
Although a step is visible at the matrix-precipitate interphase boundary, the precipitate is not sheared, but the dislocation forming the Orowan loop is absorbed into the disordered boundary. 
As expected for Orowan looping, the critical resolved shear stress to pass the precipitates is not significantly influenced by changes in the matrix-precipitate orientation relationship or precipitate shape, as well as temperature, and is well-described by the Bacon, Kocks and Scattergood model.
However, changing the type of obstacle by either removing all atoms belonging to the \MgAl precipitate and creating a void, or by fixing all their atomic positions and creating a stiff impenetrable obstacle changes the obstacle strength as well as the stress necessary for a second dislocation to pass the obstacles. 
The observed absorption of dislocations into the precipitate-matrix interface has important consequences for particle hardening, as this influences the dislocation line energy and stress field, and the absorbed Orowan loops are not available for subsequent combined cross-slip Orowan mechanisms.
Inclusion of these effects might be important for quantitative mesoscale models of particle hardening.
%
%
%
%
%
%
%
%
%
%
%
%




\section*{Acknowledgements}
The  authors acknowledge financial support by the Deutsche Forschungsgemeinschaft (DFG) 
within the Cluster of Excellence ‘‘Engineering of Advanced Materials’’ (Project EXC 315, including Bridge Funding). AV thanks Peter Felfer for his help with the figures. Computing resources were provided by the Regionales RechenZentrum Erlangen (RRZE)

\section*{References}
\bibliography{cleanedUp_bibFile.bib}

\begin{thebibliography}{10}
\expandafter\ifx\csname url\endcsname\relax
  \def\url#1{\texttt{#1}}\fi
\expandafter\ifx\csname urlprefix\endcsname\relax\def\urlprefix{URL }\fi
\expandafter\ifx\csname href\endcsname\relax
  \def\href#1#2{#2} \def\path#1{#1}\fi

\bibitem{Nembach1997}
E.~Nembach, {Particle strengthening of metals and alloys}, A Wiley-Interscience
  publication, J. Wiley, 1997.

\bibitem{Bitzek2005}
E.~Bitzek, P.~Gumbsch, {Dynamic aspects of dislocation motion: Atomistic
  simulations}, Materials Science and Engineering A 400-401~(1-2 SUPPL.) (2005)
  40--44.
\newblock \href {http://dx.doi.org/10.1016/j.msea.2005.03.047}
  {\path{doi:10.1016/j.msea.2005.03.047}}.

\bibitem{Bacon2009DIS}
D.~J. Bacon, Y.~N. Osetsky, D.~Rodney, {Chapter 88 Dislocation–Obstacle
  Interactions at the Atomic Level}, in: J.~P. Hirth, L.~B. T. D. i.~S. Kubin
  (Eds.), Dislocations in Solids, Vol.~15, Elsevier, 2009, pp. 1--90.
\newblock \href {http://arxiv.org/abs/arXiv:1011.1669v3}
  {\path{arXiv:arXiv:1011.1669v3}}, \href
  {http://dx.doi.org/https://doi.org/10.1016/S1572-4859(09)01501-0}
  {\path{doi:https://doi.org/10.1016/S1572-4859(09)01501-0}}.

\bibitem{Osetsky2003}
Y.~Osetsky, D.~Bacon, {Void and precipitate strengthening in $\alpha$-iron:
  what can we learn from atomic-level modelling?}, Journal of Nuclear Materials
  323~(2-3) (2003) 268--280.
\newblock \href {http://dx.doi.org/10.1016/j.jnucmat.2003.08.028}
  {\path{doi:10.1016/j.jnucmat.2003.08.028}}.

\bibitem{Shin2003}
C.~S. Shin, M.~C. Fivel, M.~Verdier, K.~H. Oh, {Dislocation-impenetrable
  precipitate interaction: A three-dimensional discrete dislocation dynamics
  analysis}, Philosophical Magazine 83~(31-34) (2003) 3691--3704.
\newblock \href {http://dx.doi.org/10.1080/14786430310001599379}
  {\path{doi:10.1080/14786430310001599379}}.

\bibitem{Takahashi2008}
A.~Takahashi, N.~Ghoneim, {A computational method for dislocation–precipitate
  interaction}, Journal of the Mechanics and Physics of Solids 56~(4) (2008)
  1534--1553.
\newblock \href {http://dx.doi.org/10.1016/j.jmps.2007.08.002}
  {\path{doi:10.1016/j.jmps.2007.08.002}}.

\bibitem{Kohler2005}
C.~Kohler, P.~Kizler, S.~Schmauder, {Atomistic simulation of precipitation
  hardening in $\alpha$-iron: Influence of precipitate shape and chemical
  composition}, Modelling and Simulation in Materials Science and Engineering
  13~(1) (2005) 35--45.
\newblock \href {http://dx.doi.org/10.1088/0965-0393/13/1/003}
  {\path{doi:10.1088/0965-0393/13/1/003}}.

\bibitem{Terentyev2008}
D.~A. Terentyev, G.~Bonny, L.~Malerba, {Strengthening due to coherent Cr
  precipitates in Fe-Cr alloys: Atomistic simulations and theoretical models},
  Acta Materialia 56~(13) (2008) 3229--3235.
\newblock \href {http://dx.doi.org/10.1016/j.actamat.2008.03.004}
  {\path{doi:10.1016/j.actamat.2008.03.004}}.

\bibitem{Terentyev2012}
D.~Terentyev, L.~Malerba, {Interaction of a screw dislocation with
  Cu-precipitates, nanovoids and Cu-vacancy clusters in BCC iron}, Journal of
  Nuclear Materials 421~(1-3) (2012) 32--38.
\newblock \href {http://dx.doi.org/10.1016/j.jnucmat.2011.11.037}
  {\path{doi:10.1016/j.jnucmat.2011.11.037}}.

\bibitem{Takahashi2010}
a.~Takahashi, K.~Kurata, {Dislocation dynamics based modelling of
  dislocation-precipitate interactions in bcc metals}, IOP Conference Series:
  Materials Science and Engineering 10 (2010) 012081.
\newblock \href {http://dx.doi.org/10.1088/1757-899X/10/1/012081}
  {\path{doi:10.1088/1757-899X/10/1/012081}}.

\bibitem{Polmear1994}
I.~J. Polmear, {Magnesium alloys and applications}, Materials Science and
  Technology 10~(1) (1994) 1--16.
\newblock \href {http://dx.doi.org/10.1179/mst.1994.10.1.1}
  {\path{doi:10.1179/mst.1994.10.1.1}}.

\bibitem{Luo1999}
A.~Luo, M.~O. Pekguleryuz, {Cast magnesium alloys for elevated temperature
  applications}, Journal of Materials Science 29~(20) (1999) 5259--5271.
\newblock \href {http://dx.doi.org/10.1007/BF01171534}
  {\path{doi:10.1007/BF01171534}}.

\bibitem{Hanko2002}
G.~Hanko, H.~Antrekowitsch, P.~Ebner, {Recycling automotive magnesium scrap},
  Jom 54~(2) (2002) 51--54.
\newblock \href {http://dx.doi.org/10.1007/BF02701075}
  {\path{doi:10.1007/BF02701075}}.

\bibitem{Bohlen07MSF}
J.~Bohlen, D.~Letzig, K.~U. Kainer, New perspectives for wrought magnesium
  alloys, in: Materials science forum, Vol. 546, Trans Tech Publ, 2007, pp.
  1--10.

\bibitem{Bamberger2008}
M.~Bamberger, G.~Dehm, {Trends in the Development of New Mg Alloys}, Annual
  Review of Materials Research 38~(1) (2008) 505--533.
\newblock \href {http://dx.doi.org/10.1146/annurev.matsci.020408.133717}
  {\path{doi:10.1146/annurev.matsci.020408.133717}}.

\bibitem{Wang2008b}
N.~Wang, W.-Y. Yu, B.-Y. Tang, L.-M. Peng, W.-J. Ding, {Structural and
  mechanical properties of $\text{Mg}_{17}\text{Al}_{12}$ and
  $\text{Mg}_{24}\text{Y}_{5}$ from first-principles calculations}, Journal of
  Physics D: Applied Physics 41~(19) (2008) 195408.
\newblock \href {http://dx.doi.org/10.1088/0022-3727/41/19/195408}
  {\path{doi:10.1088/0022-3727/41/19/195408}}.

\bibitem{Clark1968}
J.~Clark, {Age hardening in a Mg-9 wt.{\%} Al alloy}, Acta Metallurgica 16~(2)
  (1968) 141--152.
\newblock \href {http://dx.doi.org/10.1016/0001-6160(68)90109-0}
  {\path{doi:10.1016/0001-6160(68)90109-0}}.

\bibitem{Burssik2002}
J.~Burs{\v{s}}{\'{i}}k, M.~Svoboda, {A HREM and Analytical STEM Study of
  Precipitates in an {AZ91} Magnesium Alloy}, Microchimica Acta 139~(1-4)
  (2002) 39--42.
\newblock \href {http://dx.doi.org/10.1007/s006040200036}
  {\path{doi:10.1007/s006040200036}}.

\bibitem{Humphreys1970}
F.~Humphreys, P.~B. Hirsch, The deformation of single crystals of copper and
  copper-zinc alloys containing alumina particles-ii. microstructure and
  dislocation-particle interactions, Proc. R. Soc. Lond. A 318~(1532) (1970)
  73--92.

\bibitem{Hirsch1957}
P.~B. Hirsch, {The interpretation of the slip pattern in terms of dislocation
  movements}, J. Inst. Met 86~(13) (1957) 1958.

\bibitem{Delmas2003}
F.~Delmas, M.~Vivas, P.~Lours, M.~J. Casanove, A.~Couret, A.~Coujou, {Straining
  mechanisms in aluminium alloy 6056. In-situ investigation by transmission
  electron microscopy}, Materials Science and Engineering A 340~(1-2) (2003)
  286--291.
\newblock \href {http://dx.doi.org/10.1016/S0921-5093(02)00184-3}
  {\path{doi:10.1016/S0921-5093(02)00184-3}}.

\bibitem{Liu2011a}
G.~Liu, I.~Robertson, {Three-dimensional visualization of
  dislocation-precipitate interactions in a Al–4Mg–0.3Sc alloy using
  weak-beam dark-field electron tomography}, Journal of Materials Research
  26~(04) (2011) 514--522.
\newblock \href {http://dx.doi.org/10.1557/jmr.2010.83}
  {\path{doi:10.1557/jmr.2010.83}}.

\bibitem{Takahashi2012}
J.~Takahashi, K.~Kawakami, Y.~Kobayashi, {Consideration of
  particle-strengthening mechanism of copper-precipitation-strengthened steels
  by atom probe tomography analysis}, Materials Science and Engineering A 535
  (2012) 144--152.
\newblock \href {http://dx.doi.org/10.1016/j.msea.2011.12.056}
  {\path{doi:10.1016/j.msea.2011.12.056}}.

\bibitem{Nembach1983}
E.~Nembach, {Precipitation hardening caused by a difference in shear modulus
  between particle and matrix}, physica status solidi (a) 78~(2) (1983)
  571--581.
\newblock \href {http://dx.doi.org/10.1002/pssa.2210780223}
  {\path{doi:10.1002/pssa.2210780223}}.

\bibitem{Brown1971}
L.~M. Brown, W.~M. Stobbs, {The work-hardening of copper-silica i. A model
  based on internal stresses, with no plastic relaxation}, Philosophical
  Magazine 23~(185) (1971) 1201--1233.
\newblock \href {http://dx.doi.org/10.1080/14786437108217406}
  {\path{doi:10.1080/14786437108217406}}.

\bibitem{Hazzledine1974}
P.~M. Hazzledine, P.~B. Hirsch, {A coplanar Orowan loops model for dispersion
  hardening}, Philosophical Magazine 30~(6) (1974) 1331--1351.
\newblock \href {http://dx.doi.org/10.1080/14786437408207286}
  {\path{doi:10.1080/14786437408207286}}.

\bibitem{Monnet2011}
G.~Monnet, S.~Naamane, B.~Devincre, {Orowan strengthening at low temperatures
  in bcc materials studied by dislocation dynamics simulations}, Acta
  Materialia 59~(2) (2011) 451--461.
\newblock \href {http://dx.doi.org/10.1016/j.actamat.2010.09.039}
  {\path{doi:10.1016/j.actamat.2010.09.039}}.

\bibitem{Xiang2004}
Y.~Xiang, D.~J. Srolovitz, L.~T. Cheng, W.~E, {Level set simulations of
  dislocation-particle bypass mechanisms}, Acta Materialia 52~(7) (2004)
  1745--1760.
\newblock \href {http://dx.doi.org/10.1016/j.actamat.2003.12.016}
  {\path{doi:10.1016/j.actamat.2003.12.016}}.

\bibitem{Queyreau2010}
S.~Queyreau, G.~Monnet, B.~Devincre, {Orowan strengthening and forest hardening
  superposition examined by dislocation dynamics simulations}, Acta Materialia
  58~(17) (2010) 5586--5595.
\newblock \href {http://arxiv.org/abs/1106.3789} {\path{arXiv:1106.3789}},
  \href {http://dx.doi.org/10.1016/j.actamat.2010.06.028}
  {\path{doi:10.1016/j.actamat.2010.06.028}}.

\bibitem{Xiang2006}
Y.~Xiang, D.~J. Srolovitz, {Dislocation climb effects on particle bypass
  mechanisms}, Philosophical Magazine 86~(25-26) (2006) 3937--3957.
\newblock \href {http://dx.doi.org/10.1080/14786430600575427}
  {\path{doi:10.1080/14786430600575427}}.

\bibitem{Hatano2006}
T.~Hatano, {Dynamics of a dislocation bypassing an impenetrable precipitate:
  The Hirsch mechanism revisited}, Physical Review B - Condensed Matter and
  Materials Physics 74~(2) (2006) 1--4.
\newblock \href {http://arxiv.org/abs/0511358v2} {\path{arXiv:0511358v2}},
  \href {http://dx.doi.org/10.1103/PhysRevB.74.020102}
  {\path{doi:10.1103/PhysRevB.74.020102}}.

\bibitem{Takahashi2008a}
A.~Takahashi, Y.~Aoki, M.~Kikuchi, {Molecular dynamics simulation of
  interaction between screw dislocation and copper precipitate in iron}, Adv.
  Mater. Res. 33~(112) (2008) 895--900.
\newblock \href {http://dx.doi.org/10.4028/www.scientific.net/AMR.33-37.895}
  {\path{doi:10.4028/www.scientific.net/AMR.33-37.895}}.

\bibitem{Bacon2009}
D.~J. Bacon, Y.~N. Osetsky, {Mechanisms of hardening due to copper precipitates
  in $\alpha$-iron}, Philosophical Magazine 89~(34-36) (2009) 3333--3349.
\newblock \href {http://dx.doi.org/10.1080/14786430903271377}
  {\path{doi:10.1080/14786430903271377}}.

\bibitem{Proville2010}
L.~Proville, B.~Bak{\'{o}}, {Dislocation depinning from ordered nanophases in a
  model fcc crystal: From cutting mechanism to Orowan looping}, Acta Materialia
  58~(17) (2010) 5565--5571.
\newblock \href {http://dx.doi.org/10.1016/j.actamat.2010.06.018}
  {\path{doi:10.1016/j.actamat.2010.06.018}}.

\bibitem{Terentyev2011}
D.~Terentyev, L.~Malerba, G.~Bonny, A.~T. Al-Motasem, M.~Posselt, {Interaction
  of an edge dislocation with Cu-Ni-vacancy clusters in bcc iron}, Journal of
  Nuclear Materials 419~(1-3) (2011) 134--139.
\newblock \href {http://dx.doi.org/10.1016/j.jnucmat.2011.08.021}
  {\path{doi:10.1016/j.jnucmat.2011.08.021}}.

\bibitem{Prakash15AM}
A.~Prakash, J.~Gu\'enol\'e, J.~Wang, J.~M\"uller, E.~Spiecker, M.~Mills,
  I.~Povstugar, P.~Choi, D.~Raabe, E.~Bitzek, Atom probe informed simulations
  of dislocation–precipitate interactions reveal the importance of local
  interface curvature, Acta Materialia 92~(0) (2015) 33--45.
\newblock \href {http://dx.doi.org/10.1016/j.actamat.2015.03.050}
  {\path{doi:10.1016/j.actamat.2015.03.050}}.

\bibitem{Groh2014}
S.~Groh, {Transformation of shear loop into prismatic loops during bypass of an
  array of impenetrable particles by edge dislocations}, Materials Science and
  Engineering A 618 (2014) 29--36.
\newblock \href {http://dx.doi.org/10.1016/j.msea.2014.08.079}
  {\path{doi:10.1016/j.msea.2014.08.079}}.

\bibitem{fan2018effect}
H.~Fan, Y.~Zhu, Q.~Wang, Effect of precipitate orientation on the twinning
  deformation in magnesium alloys, Computational Materials Science 155 (2018)
  378--382.

\bibitem{fan2018precipitation}
H.~Fan, Y.~Zhu, J.~A. El-Awady, D.~Raabe, Precipitation hardening effects on
  extension twinning in magnesium alloys, International Journal of Plasticity
  106 (2018) 186--202.

\bibitem{Liao2014}
M.~Liao, B.~Li, M.~F. Horstemeyer, {Interaction Between Basal Slip and a
  $\text{Mg}_{17}\text{Al}_{12}$ Precipitate in Magnesium}, Metallurgical and
  Materials Transactions A 45~(8) (2014) 3661--3669.
\newblock \href {http://dx.doi.org/10.1007/s11661-014-2284-3}
  {\path{doi:10.1007/s11661-014-2284-3}}.

\bibitem{Liao2013}
M.~Liao, B.~Li, M.~F. Horstemeyer, {Interaction between prismatic slip and a
  $\text{Mg}_{17}\text{Al}_{12}$ precipitate in magnesium}, Computational
  Materials Science 79 (2013) 534--539.
\newblock \href {http://dx.doi.org/10.1016/j.commatsci.2013.07.016}
  {\path{doi:10.1016/j.commatsci.2013.07.016}}.

\bibitem{Liao2013a}
M.~Liao, B.~Li, M.~F. Horstemeyer, {Unstable dissociation of a prismatic
  dislocation in magnesium}, Scripta Materialia 69~(3) (2013) 246--249.
\newblock \href {http://dx.doi.org/10.1016/j.scriptamat.2013.04.008}
  {\path{doi:10.1016/j.scriptamat.2013.04.008}}.

\bibitem{Jelinek2012a}
B.~Jelinek, S.~Groh, M.~F. Horstemeyer, J.~Houze, S.~G. Kim, G.~J. Wagner,
  A.~Moitra, M.~I. Baskes, {Modified embedded atom method potential for Al, Si,
  Mg, Cu, and Fe alloys}, Physical Review B - Condensed Matter and Materials
  Physics 85~(24).
\newblock \href {http://dx.doi.org/10.1103/PhysRevB.85.245102}
  {\path{doi:10.1103/PhysRevB.85.245102}}.

\bibitem{Moitra2017}
A.~Moitra, J.~LLorca, {Atomistic simulations of dislocation/precipitation
  interactions in Mg-Al alloys and implications for precipitation hardening},
  {ArXiv e-prints}\href {http://arxiv.org/abs/1704.03487}
  {\path{arXiv:1704.03487}}.

\bibitem{Pasianot1992a}
R.~Pasianot, E.~J. Savino, {Embedded-atom-method interatomic potentials for hcp
  metals}, Physical Review B 45~(22) (1992) 12704--12710.
\newblock \href {http://dx.doi.org/10.1103/PhysRevB.45.12704}
  {\path{doi:10.1103/PhysRevB.45.12704}}.

\bibitem{Plimpton95JCP}
S.~Plimpton, Fast parallel algorithms for short-range molecular dynamics,
  Journal of Computational Physics 117 (1995) 1--19.
\newblock \href {http://dx.doi.org/10.1006/jcph.1995.1039}
  {\path{doi:10.1006/jcph.1995.1039}}.

\bibitem{Kim2009}
Y.-M.~M. Kim, N.~J. Kim, B.-J.~J. Lee, {Atomistic Modeling of pure Mg and Mg-Al
  systems}, Calphad 33~(4) (2009) 650--657.
\newblock \href {http://dx.doi.org/10.1016/j.calphad.2009.07.004}
  {\path{doi:10.1016/j.calphad.2009.07.004}}.

\bibitem{Rodney2000}
D.~Rodney, G.~Martin, {Dislocation pinning by glissile interstitial loops in a
  nickel crystal: A molecular-dynamics study}, Physical Review B 61~(13) (2000)
  8714--8725.
\newblock \href {http://dx.doi.org/10.1103/PhysRevB.61.8714}
  {\path{doi:10.1103/PhysRevB.61.8714}}.

\bibitem{Zhang2005a}
M.~X. Zhang, P.~M. Kelly, {Edge-to-edge matching and its applications: Part II.
  Application to Mg-Al, Mg-Y and Mg-Mn alloys}, Acta Materialia 53~(4) (2005)
  1085--1096.
\newblock \href {http://dx.doi.org/10.1016/j.actamat.2004.11.005}
  {\path{doi:10.1016/j.actamat.2004.11.005}}.

\bibitem{Xiao2013}
W.~Xiao, X.~Zhang, W.~T. Geng, G.~Lu, {Atomistic study of plastic deformation
  in Mg-Al alloys}, Materials Science and Engineering A 586 (2013) 245--252.
\newblock \href {http://dx.doi.org/10.1016/j.msea.2013.07.093}
  {\path{doi:10.1016/j.msea.2013.07.093}}.

\bibitem{Hagihara2018}
K.~Hagihara, K.~Hayakawa, Plastic deformation behavior and operative slip
  systems in $\text{Mg}_{17}\text{Al}_{12}$ single crystals, Materials Science
  and Engineering: A 737 (2018) 393--400.

\bibitem{Szajewski2015}
B.~A. Szajewski, W.~A. Curtin, {Analysis of spurious image forces in atomistic
  simulations of dislocations}, Modelling and Simulation in Materials Science
  and Engineering 23~(2) (2015) 025008.
\newblock \href {http://dx.doi.org/10.1088/0965-0393/23/2/025008}
  {\path{doi:10.1088/0965-0393/23/2/025008}}.

\bibitem{Sheppard2008}
D.~Sheppard, R.~Terrell, G.~Henkelman, {Optimization methods for finding
  minimum energy paths.}, The Journal of chemical physics 128~(13) (2008)
  134106.
\newblock \href {http://dx.doi.org/10.1063/1.2841941}
  {\path{doi:10.1063/1.2841941}}.

\bibitem{Guenole19FIRE}
J.~Guénolé, A.~Vaid, F.~Houllé, Z.~Xie, W.~G. Nöhring, A.~Prakash,
  E.~Bitzek, Assessment and optimization of the fast inertial relaxation engine
  (fire) for energy minimization in atomistic simulations and its
  implementation in lammps, to be published.

\bibitem{Bitzek2006}
E.~Bitzek, P.~Koskinen, F.~G{\"{a}}hler, M.~Moseler, P.~Gumbsch, M.~Moseler,
  P.~Gumbsch, {Structural Relaxation Made Simple}, Physical Review Letters
  97~(17) (2006) 170201.
\newblock \href {http://dx.doi.org/10.1103/PhysRevLett.97.170201}
  {\path{doi:10.1103/PhysRevLett.97.170201}}.

\bibitem{Nose1984a}
S.~S. Nos{\'{e}}, {A molecular dynamics method for simulations in the canonical
  ensemble}, Molecular Physics 52~(2) (1984) 255--268.
\newblock \href {http://dx.doi.org/10.1080/00268978400101201}
  {\path{doi:10.1080/00268978400101201}}.

\bibitem{Hoover1985}
H.~W. G., W.~G. Hoover, {Canonical Dynamics: Equilibrium Phase-Space
  Distributions}, Phys. Rev. A: At., Mol., Opt. Phys. 31~(3) (1985) 1695.
\newblock \href {http://dx.doi.org/10.1103/PhysRevA.31.1695}
  {\path{doi:10.1103/PhysRevA.31.1695}}.

\bibitem{Hoover1986}
W.~G. Hoover, {Constant-pressure equations of motion}, Physical Review A 34~(3)
  (1986) 2499--2500.
\newblock \href {http://arxiv.org/abs/arXiv:1011.1669v3}
  {\path{arXiv:arXiv:1011.1669v3}}, \href
  {http://dx.doi.org/10.1103/PhysRevA.34.2499}
  {\path{doi:10.1103/PhysRevA.34.2499}}.

\bibitem{Stukowski2009}
A.~Stukowski, {Visualization and analysis of atomistic simulation data with
  OVITO-the Open Visualization Tool}, Modelling and Simulation in Materials
  Science and Engineering 18~(1) (2010) 015012.
\newblock \href {http://dx.doi.org/10.1088/0965-0393/18/1/015012}
  {\path{doi:10.1088/0965-0393/18/1/015012}}.

\bibitem{Faken1994}
D.~Faken, H.~J{\'{o}}nsson, {Systematic analysis of local atomic structure
  combined with 3D computer graphics}, Computational Materials Science 2~(2)
  (1994) 279--286.
\newblock \href {http://dx.doi.org/10.1016/0927-0256(94)90109-0}
  {\path{doi:10.1016/0927-0256(94)90109-0}}.

\bibitem{Honeycutt1987}
J.~D. Honeycutt, H.~C. Andemen, {Molecular Dynamics Study of Melting and
  Freezing of Small Lennard- Jones Clusters}, Journal of Physical Chemistry
  91~(24) (1987) 4950--4963.
\newblock \href {http://dx.doi.org/10.1021/j100303a014}
  {\path{doi:10.1021/j100303a014}}.

\bibitem{Thompson09JCP}
A.~P. Thompson, S.~J. Plimpton, W.~Mattson, General formulation of pressure and
  stress tensor for arbitrary many-body interaction potentials under periodic
  boundary conditions, The Journal of Chemical Physics 131~(15) (2009) 154107.
\newblock \href {http://dx.doi.org/10.1063/1.3245303}
  {\path{doi:10.1063/1.3245303}}.

\bibitem{Zhou2003}
M.~Zhou, {A new look at the atomic level virial stress: on continuum-molecular
  system equivalence}, Proceedings of the Royal Society A: Mathematical,
  Physical and Engineering Sciences 459~(2037) (2003) 2347--2392.
\newblock \href {http://dx.doi.org/10.1098/rspa.2003.1127}
  {\path{doi:10.1098/rspa.2003.1127}}.

\bibitem{Allen1989}
M.~P. Allen, D.~J. Tildesley, {Computer Simulation of Liquids}, Oxford Science
  Publ, Clarendon Press, 1989.

\bibitem{Sutton1995}
A.~P. Sutton, Interfaces in crystalline materials, Monographs on the Physice
  and Chemistry of Materials (1995) 414--423.

\bibitem{Wolf2005}
D.~Wolf, V.~Yamakov, S.~Phillpot, A.~Mukherjee, H.~Gleiter, Deformation of
  nanocrystalline materials by molecular-dynamics simulation: relationship to
  experiments?, Acta Materialia 53~(1) (2005) 1--40.

\bibitem{Dao2007}
M.~Dao, L.~Lu, R.~Asaro, J.~T.~M. De~Hosson, E.~Ma, Toward a quantitative
  understanding of mechanical behavior of nanocrystalline metals, Acta
  Materialia 55~(12) (2007) 4041--4065.

\bibitem{Wang2011}
J.~Wang, R.~Hoagland, X.~Liu, A.~Misra, The influence of interface shear
  strength on the glide dislocation--interface interactions, Acta Materialia
  59~(8) (2011) 3164--3173.

\bibitem{Bitzek2009}
E.~Bitzek, C.~Brandl, D.~Weygand, P.~Derlet, H.~Van~Swygenhoven, Atomistic
  simulation of a dislocation shear loop interacting with grain boundaries in
  nanocrystalline aluminium, Modelling and Simulation in Materials Science and
  Engineering 17~(5) (2009) 055008.

\bibitem{Wang2003}
R.~Wang, A.~Eliezer, E.~Gutman, Microstructures and dislocations in the
  stressed {AZ91D} magnesium alloys, Materials Science and Engineering: A
  344~(1-2) (2003) 279--287.

\bibitem{Hutchinson2005}
C.~R. Hutchinson, J.-F. Nie, S.~Gorsse, Modeling the precipitation processes
  and strengthening mechanisms in a mg-al-(zn) az91 alloy, Metallurgical and
  Materials Transactions A 36~(8) (2005) 2093--2105.

\bibitem{Hirth1982}
J.~P. Hirth, J.~Lothe, {Theory of dislocations (second edition)}, second
  edition Edition, John Wiley {\&} Sons, Inc, 1982.
\newblock \href {http://dx.doi.org/10.1016/0502-8205(53)90018-5}
  {\path{doi:10.1016/0502-8205(53)90018-5}}.

\bibitem{Mathur2016}
H.~Mathur, V.~Maier-Kiener, S.~Korte-Kerzel, Deformation in the
  $\gamma$-$\text{Mg}_{17}\text{Al}_{12}$ phase at 25-278$\degree$ c, Acta
  Materialia 113 (2016) 221--229.

\bibitem{Fukuchi1975}
M.~Fukuchi, K.~Watanabe, Temperature and composition dependence of hardness,
  resistivity and thermoelectric power of the gamma-phase in the
  aluminum-magnesium system, Nippon Kinzoku Gakkaishi 39~(5) (1975) 493--498.

\bibitem{Edalati2011}
K.~Edalati, Z.~Horita, Correlations between hardness and atomic bond parameters
  of pure metals and semi-metals after processing by high-pressure torsion,
  Scripta Materialia 64~(2) (2011) 161--164.

\bibitem{Song2009}
K.~Song, H.~Fujii, K.~Nakata, {Effect of welding speed on microstructural and
  mechanical properties of friction stir welded Inconel 600}, Materials {\&}
  Design 30~(10) (2009) 3972--3978.
\newblock \href {http://dx.doi.org/10.1016/j.matdes.2009.05.033}
  {\path{doi:10.1016/j.matdes.2009.05.033}}.

\bibitem{Ragani2011}
J.~Ragani, P.~Donnadieu, C.~Tassin, J.~J. Blandin, {High-temperature
  deformation of the $\gamma$-$\text{Mg}_{17}\text{Al}_{12}$ complex metallic
  alloy}, Scripta Materialia 65~(3) (2011) 253--256.
\newblock \href {http://dx.doi.org/10.1016/j.scriptamat.2011.04.022}
  {\path{doi:10.1016/j.scriptamat.2011.04.022}}.

\bibitem{Maghsoudi2015}
M.~Maghsoudi, A.~Zarei-Hanzaki, H.~Abedi, A.~Shamsolhodaei, The evolution of
  $\gamma$-$\text{Mg}_{17}\text{Al}_{12}$ intermetallic compound during
  accumulative back extrusion and subsequent ageing treatment, Philosophical
  Magazine 95~(31) (2015) 3497--3523.

\bibitem{Kolb2013}
M.~Kolb, C.~Walther, J.~Wheeler, S.~Korte, W.~Clegg, Private communications, to
  be published.

\bibitem{Bacon1973}
D.~J. Bacon, U.~F. Kocks, R.~O. Scattergood, {The effect of dislocation
  self-interaction on the orowan stress}, Philosophical Magazine 28~(6) (1973)
  1241--1263.
\newblock \href {http://dx.doi.org/10.1080/14786437308227997}
  {\path{doi:10.1080/14786437308227997}}.

\bibitem{Scattergood1982}
R.~O. Scattergood, D.~J. Bacon, {The strengthening effect of voids}, Acta
  Metallurgica 30~(8) (1982) 1665--1677.
\newblock \href {http://dx.doi.org/10.1016/0001-6160(82)90188-2}
  {\path{doi:10.1016/0001-6160(82)90188-2}}.

\bibitem{Wang2018}
F.~Wang, B.~Li, Atomistic calculations of surface and interfacial energies of
  $\text{Mg}_{17}\text{Al}_{12}$--{Mg} system, Journal of magnesium and alloys.

\bibitem{Kelly1971}
A.~Kelly, R.~Nicholson, Strengthening methods in crystals, Materials science
  series, Applied Science, 1971.

\bibitem{Courtney2000}
H.~C. Thomas, Mechanical behavior of materials, 2000.

\bibitem{Fisher1953}
J.~Fisher, E.~W. Hart, R.~Pry, The hardening of metal crystals by precipitate
  particles, Acta Metallurgica 1~(3) (1953) 336--339.

\bibitem{Gao2002}
H.~Gao, L.~Zhang, S.~P. Baker, Dislocation core spreading at interfaces between
  metal films and amorphous substrates, Journal of the Mechanics and Physics of
  Solids 50~(10) (2002) 2169--2202.

\bibitem{Huang2018}
Z.~Huang, J.~E. Allison, A.~Misra, Interaction of glide dislocations with
  extended precipitates in mg-nd alloys, Scientific reports 8~(1) (2018) 3570.

\bibitem{Srolovitz1984}
D.~Srolovitz, M.~Luton, R.~Petkovic-Luton, D.~Barnett, W.~Nix, Diffusionally
  modified dislocation-particle elastic interactions, Acta Metallurgica 32~(7)
  (1984) 1079--1088.

\bibitem{Blum1985}
W.~Blum, B.~Reppich, B.~Wilshire, R.~Evans, Progress in creep and fracture
  (1985).

\end{thebibliography}



\section*{Supplementary material}

# Supplementary material to
## Atomistic Simulations of Basal Dislocations Interacting with $Mg_{17}Al_{12}$ Precipitates in Mg

Aviral Vaid [a], Julien Guénolé [b,a], Aruna Prakash [c,a], Sandra Korte-Kerzel [b], Erik Bitzek [a]

[a] Department of Materials Science and Engineering, Institute I, Friedrich-Alexander Universität Erlangen-Nürnberg (FAU), Martensstr. 5, 91058 Erlangen, Germany
[b] Institute of Physical Metallurgy and Metal Physics, RWTH Aachen University, Germany
[c] Micromechanical Materials Modelling (MiMM), Institute of Mechanics and Fluid Dynamics, Technische Universität Bergakademie Freiberg (TUBAF), Germany

**Table of Content**

# S1. Comparison of Interatomic Potentials

Studying the interaction of dislocations in magnesium with $Mg_{17}Al_{12}$ precipitate requires potentials that properly describe the pure Mg phase and the intermetallic phase $Mg_{17}Al_{12}$. To our knowledge, such potentials are:

- EAM potential by Liu *et. al* [1]
- MEAM potential by Jelinek *et. al* [2]
- MEAM potential by Kim *et. al* [3]
- MEAM potential by Jelinek *et. al* [4]

Pasianot and Savino had previously reported that the embedded atom method (EAM) formalism falls short in aptly describing the elastic constants of pure hcp metals [5]. Therefore, the first potential in the list above was ignored. The modified EAM (MEAM) potential by Jelinek *et. al* (2007) [2] specifically states that it is unsuitable to study the intermetallic $Mg_{17}Al_{12}$ phase and was thus also not considered.

To test the remaining potentials, several simulations were performed on model systems to compute material properties, such as lattice parameter, elastic constants, *etc.* for both Mg and $Mg_{17}Al_{12}$. For Mg, the X-, Y- and Z-axis were along $[2\bar{1}\bar{1}0]$, $[01\bar{1}0]$ and $[0001]$ respectively, while for $Mg_{17}Al_{12}$, these axes were along $[100]$, $[010]$ and $[001]$ respectively.

- The lattice parameter and cohesive energy were determined from relaxing a setup containing $5 \times 5 \times 5$ unit cells with periodic boundary conditions (PBC) in all directions.

- The elastic constants were calculated using the `Elastic` script from the LAMMPS package – the setup in this case also contained $5 \times 5 \times 5$ unit cells with PBC in all directions. The script calculates all elastic constants for a general anisotropic material by applying appropriate strains and measuring the resulting stresses. $C_{11}$, $C_{12}$, $C_{13}$, $C_{33}$, $C_{44}$ were calculated as the averages of corresponding symmetry equivalent components.
- To compute the surface energy of the basal plane of Mg, the simulation cell contained $5 \times 5 \times 5$ unit cells with fixed boundary conditions on the basal plane and PBC in the other directions. The setup was then minimized to a force norm of $10^{-8}$ eV/Å using FIRE [6]. The surface energy was computed to be $(E - N \times E_{coh})/A$, where $E$ is the total energy of the system with the surface, $N$ is the number of atoms, $E_{coh}$ is the cohesive energy of Mg and $A$ is the surface area of the basal plane.
- The vacancy formation energy of Mg was computed by removing one atom in a setup that contained $5 \times 5 \times 5$ unit cells with PBC in all directions, and to subsequently minimized it to a force norm of $10^{-8}$ eV/Å.

A comparison between the potential properties for the remaining two potentials is shown in tables below.

| Property (Mg) | Experiment/Ab-initio | Jelinek 2012 | Kim 2009 |
|---|---|---|---|
| Cohesive Energy [eV/atom] | -1.55 [7] | -1.51 | -1.55 |
| Lattice Constant [Å] | 3.209 [8] | 3.202 | 3.209 |
| c/a ratio | 1.623 [8] | 1.620 | 1.619 |
| Bulk Modulus [GPa] | 36.9 [9] | 35.71 | 36.98 |
| $C_{11}$ [GPa] | 63.5 [9] | 59.43 | 62.99 |
| $C_{12}$ [GPa] | 25.9 [9] | 23.18 | 26.03 |
| $C_{13}$ [GPa] | 21.7 [9] | 23.76 | 21.23 |
| $C_{33}$ [GPa] | 66.5 [9] | 61.18 | 69.82 |
| $C_{44}$ [GPa] | 18.4 [9] | 17.37 | 17.20 |
| Surface Energy $\gamma_{(0001)}$ [mJ/m2] | 785 [10] | 716.21 | 677.5 |
| Vacancy Formation Energy [eV] | 0.91 [11] | 0.44 | 0.89 |

| Property ($Mg_{17}Al_{12}$) | Experiment/Ab-initio | Jelinek 2012 | Kim 2009 |
|---|---|---|---|
| Enthalpy of Formation [meV/atom] | -17.0 [2] | 49.4 | -18.5 |
| Lattice Constant [Å] | 10.55 [12] | 10.73 | 10.56 |
| Bulk Modulus [GPa] | 48.3 [12] | 47.6 | 49.5 |
| $C_{44}$ [GPa] | 20.0 [12] | -2.3 | 13.6 |
| $(C_{11}-C_{12})/2$ [GPa] | 28.9 [12] | 35.4 | 25.8 |

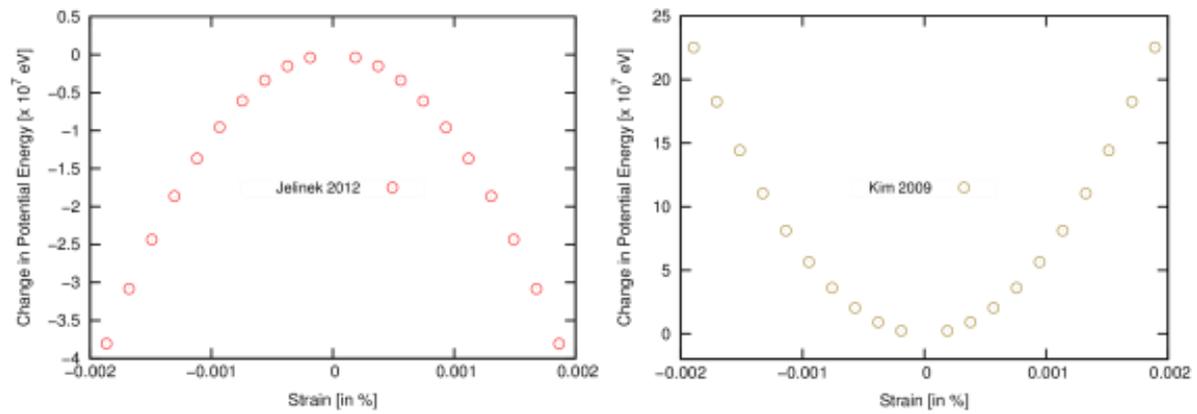

**Figure S1:** Change in potential energy of an $Mg_{17}Al_{12}$ sample with applied shear strain

The potential by Jelinek *et. al* [4] did not report any elastic constants. However, the ones computed using the methodology described above reported negative value of the elastic constant $C_{44}$ for the intermetallic $Mg_{17}Al_{12}$, which is physically unfeasible. This aberration is also shown in the fig. S1, where the change in potential energy of the setup is plotted against applied shear strain. In addition to predicting a negative elastic constant, the potential also records a positive enthalpy of formation for $Mg_{17}Al_{12}$ contrary to the experimental [13] and *ab-initio* [12] studies showing it to be a stable phase. Therefore, this potential was also excluded from our list of potentials. The potential properties of the potential by Kim *et. Al* [3] were found to be reproducible and therefore chosen as the potential of choice for further investigation in this work.

## S2. Interface termination of $Mg_{17}Al_{12}$ Embedded in Mg Matrix

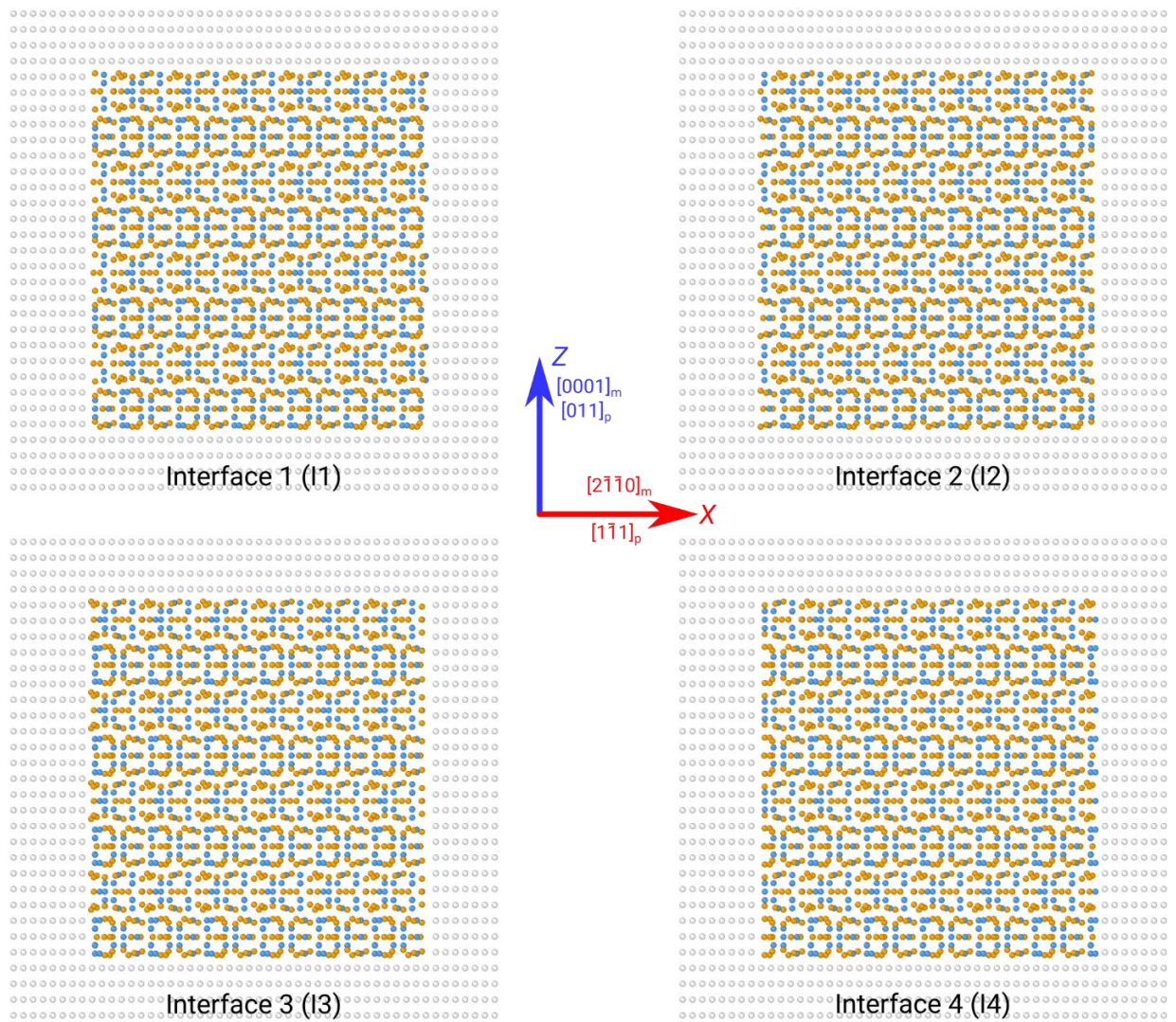

**Figure S2:** Different interface terminations of the $Mg_{17}Al_{12}$ precipitate (orange: Mg, blue Al) embedded in the Mg matrix (grey atoms) studied in the present work. The interface number correspond to those presented in the Fig. 1 of the main manuscript.

# S3. Influence of Simulation Box Size on the Critical Resolved Shear Stress

The following table shows the computed critical resolved shear stress, $\tau_C$ along with the simulation cell dimensions. No other parameters were modified other than the box aspect ratio ($L_X/L_Z$), which was shown to be the critical parameter to minimize the influence of spurious image forces [14], as compared to the *default setup*. According to Szajewski and Curtin [14], an aspect ratio of $1/\sqrt{2}$ would minimize the spurious image forces for an edge dislocation. As seen from the table below, the effect of these image forces is not significant in our simulations.

| Sample Name    | $L_X$ [Å] | $L_Y$ [Å] | $L_Z$ [Å] | ($L_X/L_Z$) | $L$ [Å] | $D$ [Å] | $\tau_C$ [MPa] |
|----------------|-----------|-----------|-----------|-------------|---------|---------|----------------|
| *Default setup*| 400       | 200       | 200       | 2           | 122     | 78      | 245 ± 5        |
| Alt. setup 1   | 500       | 200       | 200       | 2.5         | 122     | 78      | 235 ± 5        |
| Alt. setup 2   | 400       | 200       | 500       | 0.8         | 122     | 78      | 245 ± 5        |

# S4. Anisotropic Shear Modulus

The anisotropic shear modulus was calculated using the LAMMPS package – the setup in this case contained $5 \times 5 \times 5$ unit cells with periodic boundary conditions in all directions. For Mg, the X-, Y- and Z-axis were along $[2\bar{1}\bar{1}0], [01\bar{1}0]$ and $[0001]$ respectively, while for Mg17Al12, these axes were along $[1\bar{1}1], [21\bar{1}]$ and $[011]$ respectively. The elastic constants were calculated using the `Elastic` script from the LAMMPS package. The script calculates all elastic constants for a general anisotropic material by applying appropriate strains and measuring the resulting stresses. The anisotropic shear modulus was calculated as the averages of corresponding symmetry equivalent components.

$$\mu'(Mg) = 17.2 \; GPa$$

$$\mu'(Mg_{17}Al_{12}) = 17.6 \; GPa$$

## S5. Interface Width of $Mg_{17}Al_{12}$ Embedded in Mg Matrix

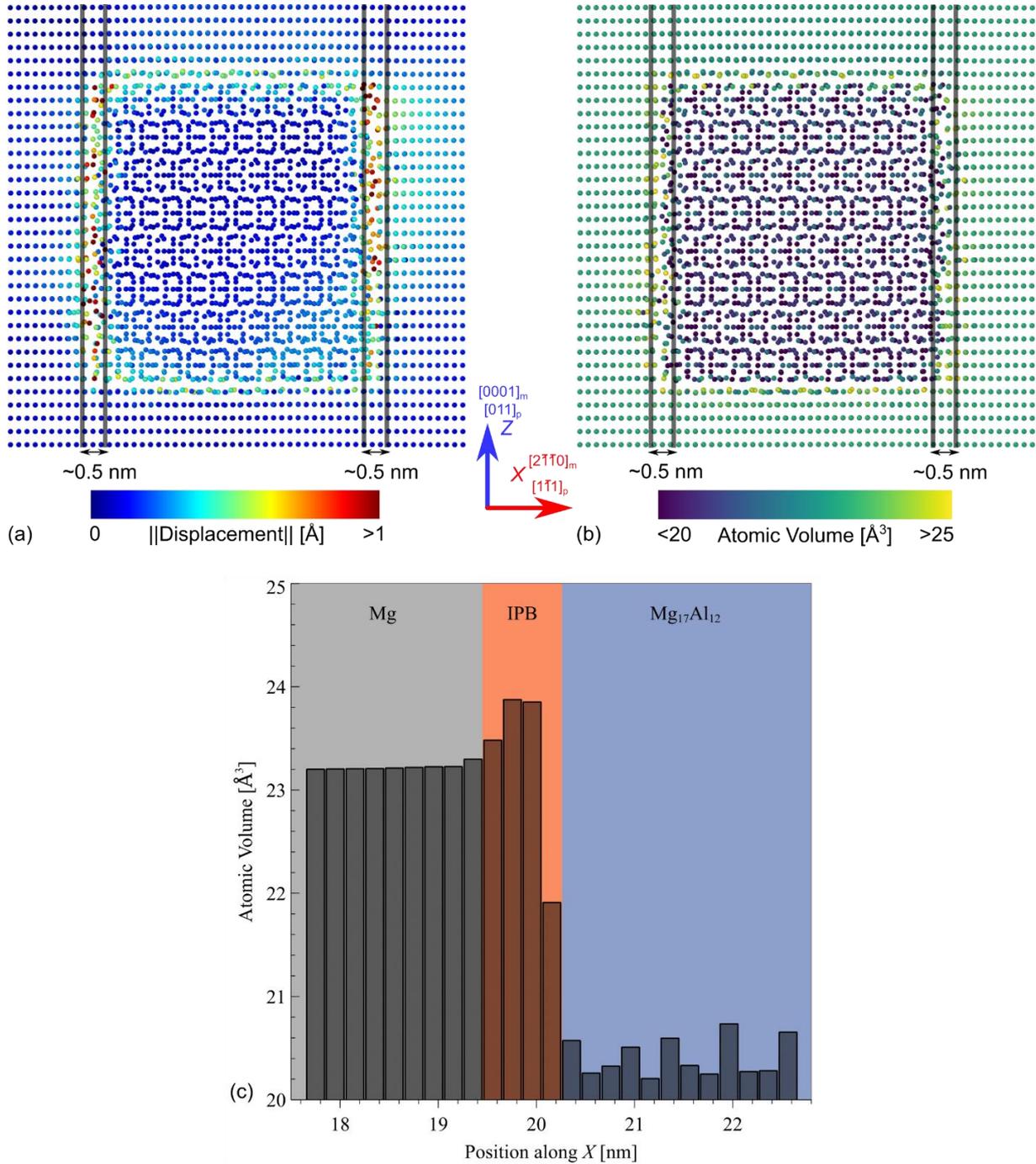

**Figure S3:** Side view showing the width of the Mg-$Mg_{17}Al_{12}$ interface for a relaxed precipitate in the matrix following the Burgers orientation relationship computed using (a) Displacement magnitude with respect to the perfect lattice positions, and (b) atomic volume. (c) Histogram of atomic volume for the region around the left interface with a binning width of 2Å showing an increased excess volume.

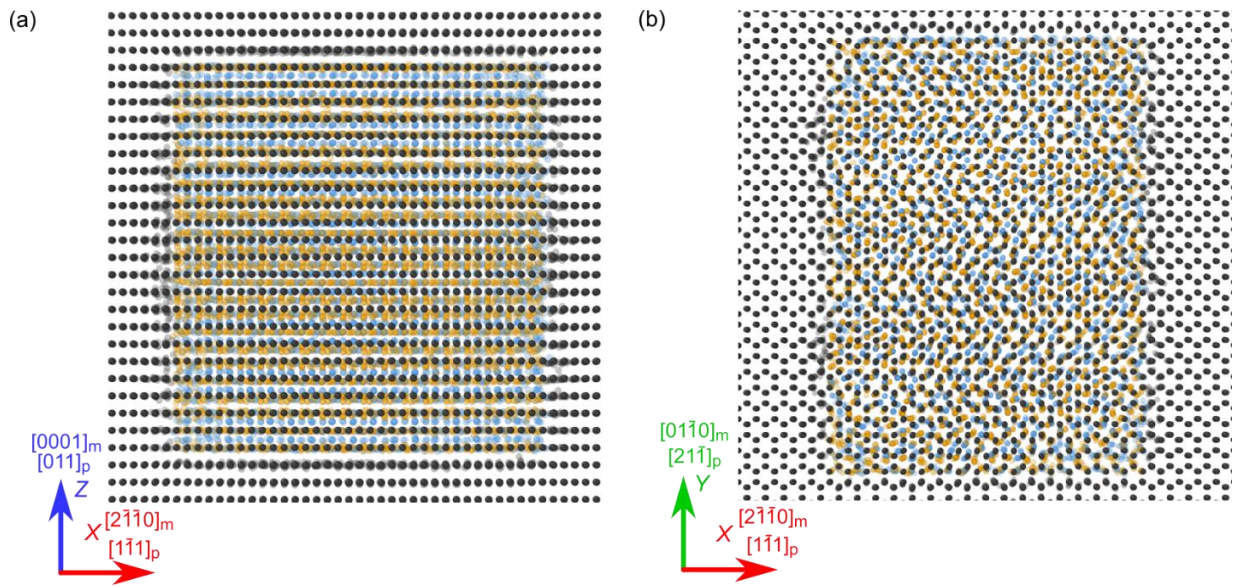

**Figure S4:** (a) Side view and (b) top view showing the local disordered region at the Mg-Mg$_{17}$Al$_{12}$ IPB for a relaxed precipitate in the matrix following the Burgers orientation relationship. The matrix atoms are dark grey, while the atoms in the precipitate are colored based on atomic species – Mg (orange), Al (blue).

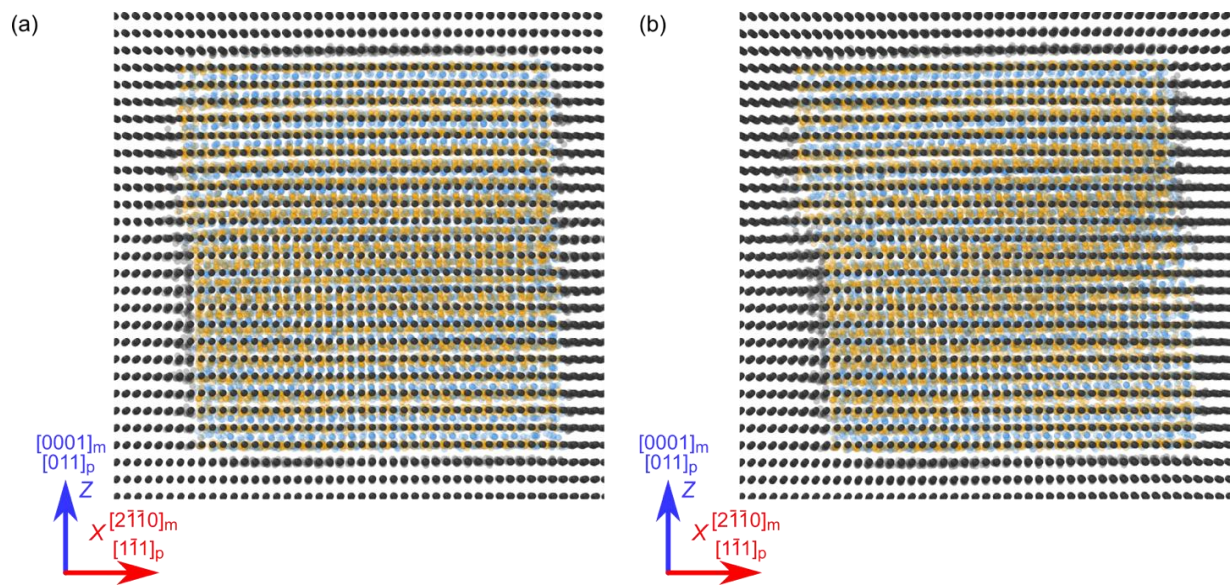

**Figure S5:** Side view showing the relaxed local disordered region at the Mg-Mg$_{17}$Al$_{12}$ IPB for the *default setup* (a) with one absorbed dislocation at $\tau = 240$ MPa, (b) with two absorbed dislocations after passage of the first dislocation at $\tau = 250$ MPa. The matrix atoms are dark grey, while the atoms in the precipitate are colored based on atomic species – Mg (orange), Al (blue).

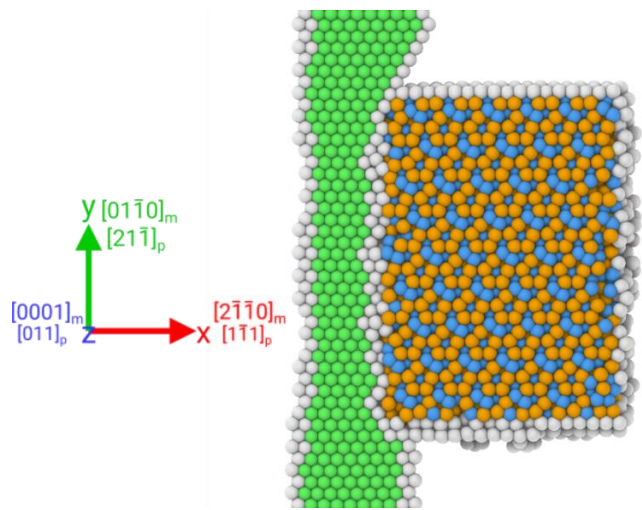

**Figure S6:** Top view of a cut just above the dislocation glide plane showing an edge dislocation in front of the precipitate upon energy minimization at $\tau = 0$ MPa for the *default setup*. The defect atoms in the Mg matrix are colored as green (fcc) and white (other), while the atoms in the precipitate are colored based on atomic species – Mg (orange), Al (blue). The hcp coordinated matrix atoms are not shown.

## S6. Dislocation Precipitate Interaction using NVT at $T = 300$ K

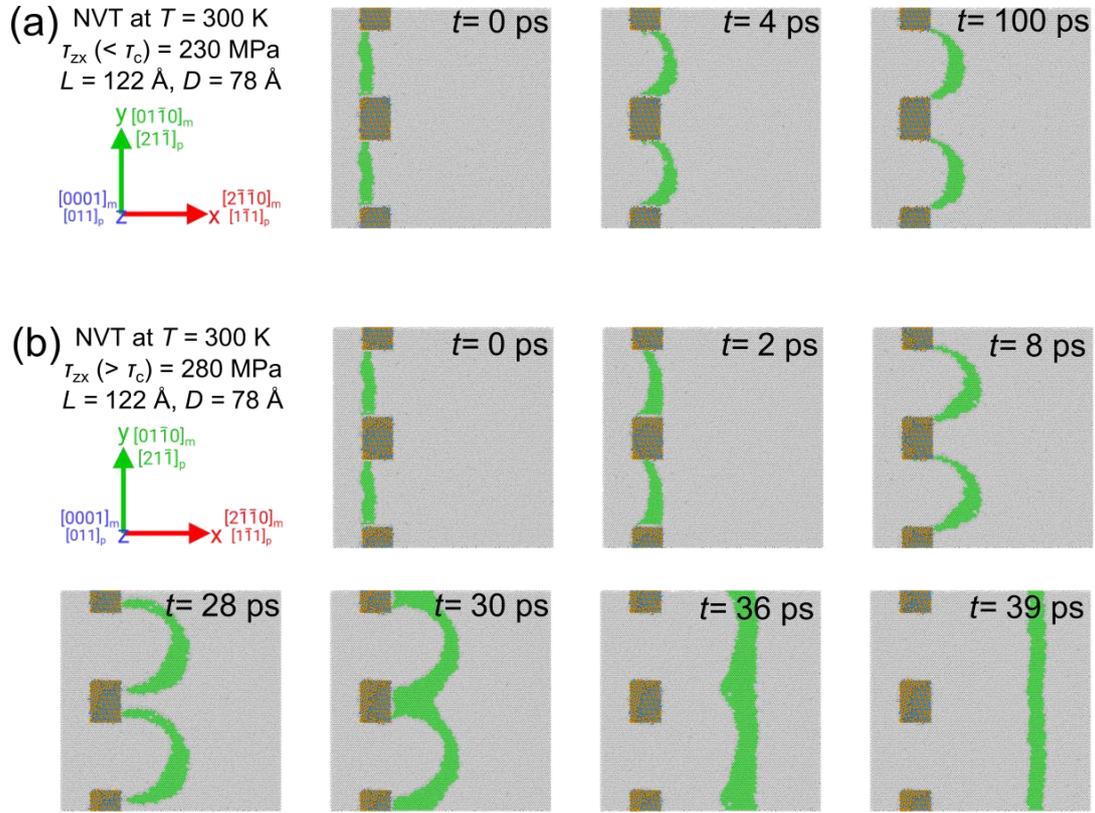

**Figure S7:** Infinite basal edge dislocation interacting with a periodic array of $Mg_{17}Al_{12}$ precipitates using molecular dynamics NVT simulation at T = 300 K with an applied shear stress (a) smaller, and (b) larger than the critical resolved shear stress ($L = 122$Å, $D = 78$Å). The Mg atoms in the matrix are colored grey and the defect fcc atoms in the Mg matrix green, while the atoms in the precipitate are colored based on atomic species – Mg (orange), Al (blue).

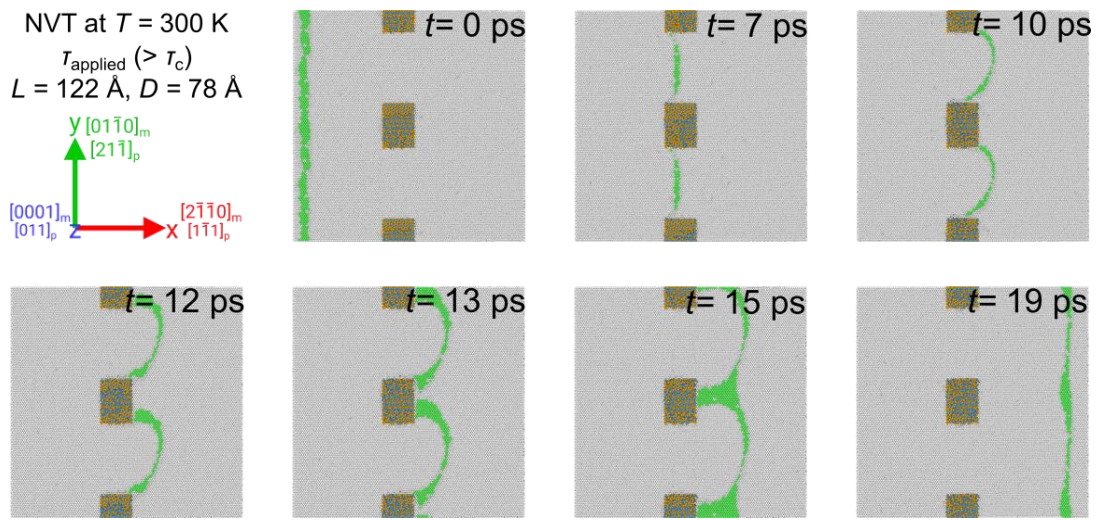

**Figure S8:** Infinite 30° mixed dislocation interacting with a periodic array of $Mg_{17}Al_{12}$ precipitates using molecular dynamics NVT simulation at T = 300 K with an applied shear stress larger than the critical resolved shear stress (L = 122Å, D = 78Å). The Mg atoms in the matrix are colored grey and the defect fcc atoms in the Mg matrix green, while the atoms in the precipitate are colored based on atomic species – Mg (orange), Al (blue).

## S7. Unloading the simulation cell

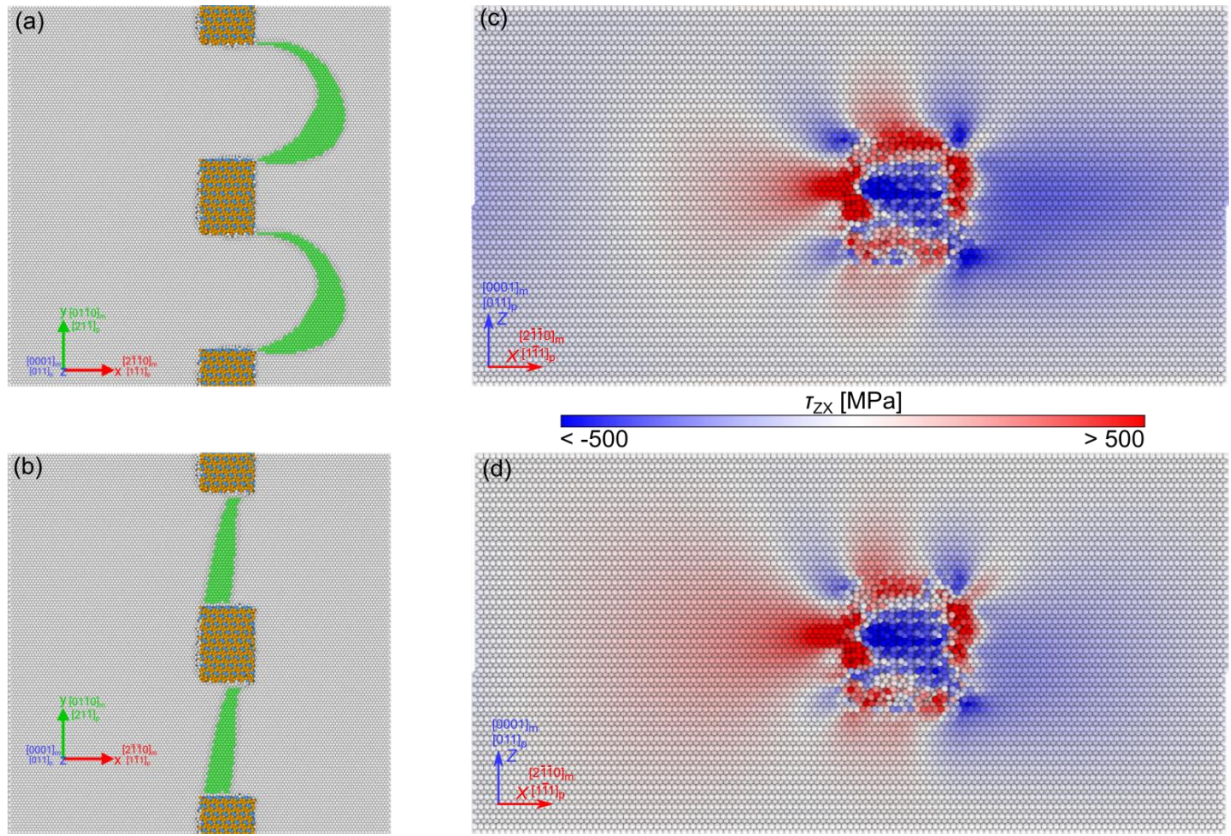

**Figure S9:** Top view of a relaxed dislocation (a) before and (b) after unloading the simulation cell for the *default* setup that was previously loaded at $\tau(<\tau_C) = 240$ MPa. (c) and (d) show the corresponding stress states from the side view. The color code for (a) and (b) is as follows: The Mg atoms in the matrix are colored grey and the defect fcc atoms in the Mg matrix green, while the atoms in the precipitate are colored based on atomic species – Mg (orange), Al (blue). The color code for the stress state is shown in the figure.

# References:

[1]     X.Y. Liu, P.P. Ohotnicky, J.B. Adams, C. Lane Rohrer, R.W. Hyland, Anisotropic surface segregation in Al-Mg alloys, Surf. Sci. 373 (1997) 357–370. doi:10.1016/S0039-6028(96)01154-5.

[2]     B. Jelinek, J. Houze, S. Kim, M.F. Horstemeyer, M.I. Baskes, S.G. Kim, Modified embedded-atom method interatomic potentials for the Mg-Al alloy system, Phys. Rev. B - Condens. Matter Mater. Phys. 75 (2007) 1–9. doi:10.1103/PhysRevB.75.054106.

[3]     Y.M. Kim, N.J. Kim, B.J. Lee, Atomistic Modeling of pure Mg and Mg-Al systems, Calphad Comput. Coupling Phase Diagrams Thermochem. 33 (2009) 650–657. doi:10.1016/j.calphad.2009.07.004.

[4]     B. Jelinek, S. Groh, M.F. Horstemeyer, J. Houze, S.G. Kim, G.J. Wagner, A. Moitra, M.I. Baskes, Modified embedded atom method potential for Al, Si, Mg, Cu, and Fe alloys, Phys. Rev. B - Condens. Matter Mater. Phys. 85 (2012). doi:10.1103/PhysRevB.85.245102.

[5]     R. Pasianot, E.J. Savino, Embedded-atom-method interatomic potentials for hcp metals, Phys. Rev. B. 45 (1992) 12704–12710. doi:10.1103/PhysRevB.45.12704.

[6]     J. Guénolé, A. Vaid, F. Houllé, Z. Xie, W.G. Nöhring, A. Prakash, E. Bitzek, Assessment and optimization of the fast inertial relaxation engine (fire) for energy minimization in atomistic simulations and its implementation in lammps, To Be Publ. (2019).

[7]     M.I. Baskes, R.A. Johnson, Modified embedded atom potentials for HCP metals, Model. Simulations Mater. Sci. Eng. 2 (1994) 147–163.

[8]     C.S. Barrett, T.B. Massalski, Structure of Metals: Crystallographic Methods, Principles And Data, Pergamon Press, Oxford, 1980.

[9]     G. Simmons, Single crystal elastic constants and calculated aggregate properties: A Handbook, SOUTHERN METHODIST UNIV DALLAS TEX, 1965.

[10]    W.R. Tyson, W.A. Miller, Surface free energies of solid metals: Estimation from liquid surface tension measurements, Surf. Sci. (1977). doi:10.1016/0039-6028(77)90442-3.

[11]    N. Chetty, M. Weinert, T.S. Rahman, J.W. Davenport, Vacancies and impurities in aluminum and magnesium, Phys. Rev. B. (1995). doi:10.1103/PhysRevB.52.6313.

[12]    N. Wang, W.Y. Yu, B.Y. Tang, L.M. Peng, W.J. Ding, Structural and mechanical properties of Mg17Al12 and Mg24Y5 from first-principles calculations, J. Phys. D. Appl. Phys. 41 (2008). doi:10.1088/0022-3727/41/19/195408.

[13]    J.A. Brown, J.A. Brown, D. Laboratories, S. Park, The Thermodynamic Properties of Solid Al-Mg Alloys, 1 (1970).

[14]    B.A. Szajewski, W.A. Curtin, Analysis of spurious image forces in atomistic simulations of dislocations, Model. Simul. Mater. Sci. Eng. 23 (2015). doi:10.1088/0965-0393/23/2/025008.




\end{document}